\documentclass[useAMS,usenatbib,babel]{mn2e}

\usepackage[english,english]{babel}
\usepackage{amsmath}
\usepackage{amssymb,amsfonts,textcomp}
\usepackage{array}
\usepackage{supertabular}
\usepackage{placeins}
\usepackage{hhline}
\usepackage[usenames]{color}
\usepackage{soul}
\usepackage{widetext}
\usepackage{textcomp}
\usepackage{graphicx}
\usepackage{dcolumn}
\usepackage{bm}
\usepackage{comment}
\usepackage{color}

\usepackage{hyperref}
 \hypersetup{dvips, colorlinks=true, linkcolor=black, citecolor=black, filecolor=black, urlcolor=black}
 \usepackage{float}
\usepackage[caption=false]{subfig}

\newcommand{\mD}{{\cal D}}
\newcommand{\mP}{{\cal P}}

\newcommand{\dd}{\hbox{d}}

\newcommand{\tree}{{\rm tree}}

\newcommand{\hx}{{\hat{x}}}

\newcommand{\mgen}{{\cal M}}
\newcommand{\mF}{{\cal F}}
\newcommand{\hrho}{{\hat\rho}}

\definecolor{Blue}{rgb}{0,0.08,0.65}
\definecolor{Green}{rgb}{0.2,0.55,0.35}
\definecolor{grey}{rgb}{0.75,0.75,0.75}
\definecolor{Orange}{rgb}{1.0,0.5,0.15}
\definecolor{brown}{rgb}{0.7,0.25,0.0}
\definecolor{Pink}{rgb}{1.0,0.5,0.5}
\definecolor{darkerred}{rgb}{0.8,0,0}
\definecolor{darkerblue}{rgb}{0,0,0.8}

\def\mod#1{#1}
\begin{document}

\title[Large-deviation statistics of the log-density]{Back in the saddle: \\ Large-deviation statistics of the cosmic log-density field}
\author[ C.~ Uhlemann, S.~Codis,   C.~Pichon,  F.~ Bernardeau and P.~Reimberg ]{
\parbox[t]{\textwidth}{
 C. Uhlemann$^{1}$,  S. Codis$^{2}$, C. Pichon$^{3}$,  F. Bernardeau$^{3,4}$ and  P.~Reimberg$^{3,4}$}
\vspace*{6pt}\\
\noindent
$^{1}$  {Institute for Theoretical Physics \& Center for Extreme Matter and Emergent Phenomena, Utrecht University,}\\
$\quad$ Leuvenlaan 4, 3584 CE Utrecht, The Netherlands\\
$^{2}$ {Canadian Institute for Theoretical Astrophysics, University of Toronto,}\\
$\quad$ 60 St. George Street, Toronto, ON M5S 3H8, Canada\\
$^{3}$ {Sorbonne Universit\'es, UPMC Univ Paris 6 et CNRS, UMR 7095, Institut d'Astrophysique de Paris,}\\
$\quad$ 98 bis bd Arago, 75014 Paris, France \\
$^{4}$ CNRS \& CEA, UMR 3681, Institut de Physique Th\'eorique, F-91191 Gif-sur-Yvette, France
}
\maketitle
\begin{abstract}
We present a first principle approach to obtain analytical predictions for spherically-averaged cosmic densities in the mildly non-linear regime that go well beyond what is usually achieved by standard perturbation theory. A large deviation principle allows us to compute the leading-order cumulants of average densities in concentric cells. In this symmetry, the spherical collapse model leads to cumulant generating functions that are robust for finite variances and free of critical points when logarithmic density transformations are implemented. They yield in turn accurate density probability distribution functions (PDFs) from a straightforward saddle-point approximation valid for all density values. Based on this easy-to-implement modification, explicit analytic formulas for the evaluation of the one- and two-cell PDF are provided. The theoretical predictions obtained for the PDFs are accurate to a few percent compared to the numerical integration, regardless of the density under consideration and in excellent agreement with N-body simulations for a wide range of densities. This formalism should prove valuable for accurately probing the quasi-linear scales of low redshift surveys for arbitrary primordial power spectra.
 \end{abstract}
 \begin{keywords}
 cosmology: theory ---
large-scale structure of Universe ---
methods: analytical, numerical 
\end{keywords}

\section{Introduction}
Given the increasing amount of data released 
by  large galaxy surveys, such as the BOSS survey \citep{boss}, DES \citep{Abbott:2005bi} and in the coming years Euclid~\citep{Euclid} and LSST \citep{lsst}, it is becoming crucial for astronomers to exploit observations of the large-scale structure of the Universe in
the best possible way. This task is  difficult as the large-scale structure of the Universe is the result of the interplay between the
cosmological parameters, such as the amount of dark matter and dark energy, the initial metric perturbations, the non-linearities in the cosmic fluid evolution, not to mention the impact of baryonic physics on  small-scales -- down to the stellar mass scale -- through the back-reaction of baryons onto the large-scale structures. Making use of the  statistical properties of the large-scale structure to extract information on fundamental cosmological parameters is therefore a daunting task.

Hence it is  crucial  to taylor complementary sets of observables that can help disentangle all of those effects.
The most commonly used tools to extract statistical information from the observed galaxy distribution are the $N$-point correlation functions \citep[e.g][]{Scoccimarro98}, which quantify how  galaxies are clustered, and primarily the two-point correlation function or its Fourier counterpart, the power spectrum. 
Observations of the cosmic microwave background strongly support the idea that the initial metric perturbations followed Gaussian statistics to an extremely good accuracy \citep{2015arXiv150201592P}. As a result, the statistical properties of the large-scale structure of the universe at its early stages, or equivalently in this context at large enough scales, are entirely characterized by this two-point correlation function. 

At  later times, when the typical density contrasts (or velocity gradients) become large,  the cosmic fluid rapidly develops  nonlinear structures.
In particular the power spectrum of the density field  evolves nontrivially via the induced mode coupling. This can be captured for instance with the help of perturbation theory (PT) approaches \citep[see][]{2002PhR...367....1B,2006PhRvD..73f3519C,2008ApJ...674..617T,2008JCAP...10..036P,2012PhRvD..86j3528T,2012arXiv1206.2926C}. Yet, the strength of these couplings, and in particular  the coupling between  small and  large scales \citep[see][]{2014PhRvD..89b3502B,2013arXiv1309.3308B,2014arXiv1411.2970N},  limits in part the relevance of such techniques. One should indeed keep in mind that for PT, the density perturbations are assumed to be small everywhere, which is obviously not the case in the present universe. 
Overcoming such a limitation is very challenging and usually relies on phenomenological methods qualified on $N$-body simulations, such as the halo models \citep{2002PhR...372....1C}.

There exists however a first principle approach based on the application of Large Deviation Theory which deals with the rate at which probabilities of certain events decay as a natural parameter of the problem decreases rapidly enough to zero. The applicability of Large Deviation Theory in this context has been discovered and partially explored in various cosmological papers over the last decades \citep[e.g ][]{Bernardeau92,1994A&A...291..697B,2000A&A...364....1B} and more precisely in \cite{Bernardeau14}, although not from the Large Deviation Principle (LDP) perspective that was recently presented in 
 \cite{LDPinLSS} in this context. The identification of a regime where LDP can be applied gives a sound framework to explore statistical properties of the fields without assuming that field fluctuations are small everywhere, the only assumption being that the typical values -- the variance -- of the quantities under consideration, for instance of the density filtered at a given scale, are small. The reach of such an approach is then potentially far wider than standard PT. The drawback is that demonstrating the applicability of the LDP is in general very complicated, and the proof is only possible in specific geometries that are special enough to allow an explicit mapping between the final configuration and the initial field values. 
Once established, the LDP implies that the probability distribution has an exponential decay that is driven by the so-called rate function. In the LSS context, the rate function can be found for highly symmetric configuration (spherical or cylindrical symmetry) for which the full 
non-linear evolution  of the dynamical equations  (the so-called spherical or cylindrical collapse model) are known exactly. 
Indeed matter contained in a spherical cell (in a static, asymptotically flat spacetime) evolves independently of the external ambient, hence  analytic solutions of the gravitational collapse can be obtained explicitly.
For corresponding observables, such as densities in concentric spheres or discs, they yield very accurate analytical predictions in the mildly non-linear regime, well beyond what is usually achieved using other estimators. 

 The aim of the paper is to extend further  the use of such an approach. In particular it will be stressed that in the LDP context, any nonlinear functions of the density in concentric cells can be considered, via the so called contraction principle as explained in \cite{LDPinLSS}, hence broadening the range of  possible applications. We will see how such choices affect
the predicted probability density functions (PDFs), and more importantly how it opens the way to having fully analytical predictions that can straightforwardly be implemented in real surveys while considering arbitrary underlying cosmic models.

A particular emphasis will be put on logarithmic transformations of the density field, which have attracted some interest in the context of cosmic structure formation since the PDF of the density field was found to be nearly log-normal \citep{hubble34, Hamilton85,1991MNRAS.248....1C,1994ApJ...435..536C,2001ApJ...561...22K}. 
In particular, the power spectrum of the log-density is known to be less prone to non-linearities \citep{2009ApJ...698L..90N,2015A&A...574A..86G} {which stimulated analytical studies within perturbation theory. In \cite{2003ApJ...583L...1S} tree-level perturbation theory in the log-density field was considered and connected to the dominant part of first-order PT in the density and higher partial loop corrections. Based on a general mapping for nonlinear bias formulated by \cite{Fry93}, especially the variance of the log-density was found to be smaller than the variance of the density.} Note however, that \cite{2011ApJ...735...32W} showed that  the log-density does not significantly enhance the validity regime of standard PT calculations based on the density. \cite{Carron11} also highlighted the limited information of log-normal statistics in the strongly non-linear regime.  In this paper, we are revisiting the use of log-densities in a different and complementary context: large deviation statistics.

The idea here is not to rely on the Gaussianity of the log-density field but on the weak dependence of its cumulants with respect to the variance of the density field. Let us make clear that here we do not expect the log-density  to arise naturally  a priori, but rather as a prime example of a nonlinear mapping that will prove particularly advantageous. The adequacy of this approach will be established by means of $N$-body simulations verifying the weak dependence of the cumulants of the log-density on the variance. This key property will indeed allow us to extend the domain of applicability of the LDP well beyond its zero variance limit. We will in particular build very accurate analytical models based on the saddle approximation for the one and two cells PDF of the density within concentric cells which match simulations up to variances of order one.

The outline of the paper is the following. We present in Section~\ref{sec:LDP} the large deviation principle that allows to obtain one-point statistics for the density in concentric spheres. We describe how to obtain the density PDF from the rate function using spherical collapse dynamics and  the saddle-point approximation, and demonstrate how the  log-density  mapping remedies technical difficulties that impeded this construction for the density. In Section~\ref{sec:cumulants} we present $N$-body measurements of the cumulants and their generating functions for both the density and the log-density in order to establish that the log-density is more resilient to changes in the variance. We formulate a mapping to relate the cumulants of the log-density to those of the density and discuss their relation based on measurements as well as analytical predictions relying on the LDP and perturbation theory, respectively. 
In Section~\ref{sec:PDFs} we finally present the analytic predictions from the LDP applied to the log-density for the one and two-cell PDF of the density, which give an excellent match to the simulations {and substantially extend the reach of the saddle-point approximation}. Section~\ref{sec:conclusion} wraps up discusses promising perspectives.

\section{The log density mapping}
\label{sec:LDP}

For the sake of clarity, let us first present  the formalism and \mod{explain how a change of variable can allow for fully analytical predictions.}
Consider one sphere ${\cal S}$ of radius $R$  centred on a given location in space. 
Our goal is to derive a working model for 
the PDF, $\mP_{R}(\hrho)$, of the density in ${\cal S}$ denoted by $\hrho$   
 and rescaled so that $\left\langle \hrho\right \rangle=1$.
In order to achieve this, we will rely here on the  LDP to connect the cumulant generating function to the PDF,
while assuming that the  variance, $\sigma^2(R)$, of the field fluctuation within that sphere is small enough.
For a complementary intuitive rather than a mathematically precise description of the connection,  see Appendix~\ref{App:Intuitive} and \cite{bcp15}.

\subsection{The Large Deviation Principle} 
\label{fb:theLDP}

Let us consider the \textsl{scaled} cumulant generating function (SCGF), $\varphi_{\hrho}(\lambda)$, defined from the cumulant generating function, $\phi_{\hrho}(\lambda)$, as
\begin{equation}
\label{def:SCGF}
\varphi_{\hrho}(\lambda)= \lim_{\sigma^{2}\to0}\sigma^{2}\phi_{\hrho}\left(\frac{\lambda}{\sigma^{2}}\right)= \lim_{\sigma^{2}\to0}\sigma^{2}
\log\left[\langle e^{\lambda \hrho/\sigma^{2}}\rangle \right]\,,
\end{equation}
where $\langle .\rangle$ stands for ensemble average, and $\sigma^{2}$ is, or scales like, the variance of $\hrho$.
The existence of a LDP for the variable $\hrho$ implies that the SCGF is given by the Legendre-Fenchel transform of the
rate function, $\psi(\hrho)$, as
\begin{equation}
\varphi_{\hrho}(\lambda)=\sup_{\hrho}\left[\lambda\hrho-\psi(\hrho)\right]\,,
\end{equation}
where the rate function is by definition the leading behaviour of the $\log$ of the density PDF,
\begin{equation}
\label{ratefct}
\psi(\hrho)=-\lim_{\sigma^{2}\to 0} \sigma^2\log \mP(\hrho).
\end{equation}

A key consequence of the LDP of relevance here is that different rate functions of functions (or functionals) of random variables related by continuous maps
are closely connected through the contraction principle. More precisely, we have 
\begin{equation}
\psi(\hrho)=\inf_{\{\tau\}\ {\rm such}\ \hrho=\mF(\{\tau\})}\ \psi_{\tau}(\{\tau\})\,, \label{eq:contract}
\end{equation}
whatever the continuous transform $\mF$ one considers (it does not have to be monotonic nor single valued). 
The infimum in equation~(\ref{eq:contract}) encodes the general idea that any large deviation follows the least unlikely of
all the unlikely  transformations. 
This transformation  can be  an active mapping (e.g. temporal evolution)  or a passive one (e.g. taking the log of the density);
we will make use of both here.

Consequences of this principle are numerous: in particular, the rate function of $\hrho$ can be computed from the  initial density field if the mapping between the initial configuration and the final field value is known, or more specifically, if one is able to identify the leading field configuration that will contribute to this infimum.

In spherically symmetric configurations, this is precisely what can be conjectured: $\hrho$ is in one-to-one correspondence with the 
linear density contrast in a cell centered on the same point and that contains the same mass.
Then, one can give the explicit expression of the rate function
\begin{equation}
\label{Psiquad}
\psi(\hrho)=\sigma^2(R)\Psi_R(\hrho) \,,\quad \Psi_R(\hrho)= \frac{1}{2\sigma^2(r)} \tau(\hrho)^2\,,
\end{equation}
where the smoothing scale $r$ is such that
$r^{3}=R^{3}\hrho$ and $\hrho=\zeta_{\rm SC}(\tau)$, with $\zeta_{\rm SC}$ the non-linear mapping between the linear overdensity within radius $r$ and the non-linear density within radius $R$ as given by the spherical collapse dynamics (see \cite{LDPinLSS} for details). The decay-rate function $\Psi_R(\hrho)$ drives the exponential decay of the PDF, as one can see from equation~\eqref{ratefct}.

Moreover, and this is the property  which is central to this paper, 
the same principle states that equation~\eqref{Psiquad} is also the rate function of any variable $\mu$, corresponding to a \textsl(monotonic) 
transformation of $\hrho$, $\mu=\mu(\hrho)$, so that $\Psi_R(\mu) = \Psi_R(\hrho(\mu))$. 
Consequently, the scaled cumulant generating function, $\varphi_{R,\mu}(\lambda)$, of such a variable is nothing
but the Legendre-Fenchel transform of the corresponding rate function, defined as
\begin{equation}
\label{phiLegendre}
\varphi_{R,\mu}(\lambda)=\sup_{\tau}\left[\lambda\mu(\tau)-\frac{\sigma^2(R)}{2\sigma^2(r)} \tau^2\right].
\end{equation}

At this stage it is important to note that the Legendre-Fenchel transform reduces to a Legendre transform provided the rate function is convex. {In that case, the Legendre transform can  be computed from simple variational calculations such that $\mu(\lambda)$ is given implicitly by the stationary condition $\lambda  =\partial \psi/\partial \mu$.} Note also that, whereas the rate function values are the same
for corresponding variables, its convexity properties depends on the choice of variables.

What are  the subsequent steps to build the density PDF? It can be obtained from an Ansatz for the cumulant generating function which
 is assumed to naturally 
 match its  asymptotic $\sigma^{2}\to 0$ limit,
\begin{equation}
\label{hyp:CGF}
\phi_{R,\hrho}(\lambda)=\frac{1}{\sigma^{2}}\varphi_{R,\hrho}(\lambda\sigma^{2})\,,
\end{equation}
so that we are now extrapolating the LDP result to  finite variances. 
The last step is to compute the inverse Laplace transform of the moment generating function, so as to write the PDF of $\hrho$ 
for a given variance $\sigma^2(R)$ as 
\begin{equation}
\label{PDFfromphi}
\mP_{R}(\hrho) = \int_{-i {\infty}}^{+i \infty} \frac{\dd\lambda}{2\pi i} \exp[-\lambda\hrho+\phi_{R,\hrho}(\lambda)] \,,
\end{equation}
where $\phi_{\hrho}(\lambda)$ is a function of $\sigma^2(R)$ given by equation~(\ref{hyp:CGF}).
Here $\sigma^2$ can actually be adjusted to match the measured variance if necessary.
The integration in equation~(\ref{PDFfromphi}) requires an analytic extension of the cumulant generating function in the complex plane, which has also been noted in \mod{\cite{Fry85,Gaztanaga00,Valageas2002a} to name just a few examples of previous work.} 

The analytical properties of this extension depend on the choice of variables. 
In \mod{previous works including} \cite{Bernardeau14} the freedom in choosing a mapping $\mu =\mu(\hrho)$ was ignored and only $\hrho$ was considered to obtain the PDF. \footnote{However, the LDP was used implicitly via the active mapping between initial and final density, see Appendix~\ref{App:Intuitive}.} This leads to a critical point along the real axis that a better choice of variable can avoid, as we precisely show now.

\subsection{Avoiding criticality by using the log-density} 

It should be clear from equation~(\ref{PDFfromphi}) 
that our ability to make predictions on the shape of the PDF depends crucially on the analytic properties of the cumulant generating functions. \mod{In particular, the existence of critical points for $\varphi_{\hrho}(\lambda)$ that arise from the Legendre tranformation of the decay-rate function $\Psi_R(\hrho)$} can make it difficult to perform the explicit integration in the complex plane.
In the low-density regime, 
the inverse Laplace transform in equation~\eqref{PDFfromphi} can in principle be computed via a saddle-point
approximation, taking advantage of the fact that the variance is small, leading to the form

\begin{align}
\label{PDFfromPsi}
\mP_{R}(\hat\rho) = \sqrt{\frac{\Psi''_R[\hrho]}{2\pi}} \exp\left(-\Psi_R[\hrho]\right)\,.
\end{align}
\mod{Note that, the saddle point approximation has been used in \cite{Fry85} in combination with a cumulant generating function from a hierarchical model rather than spherical collapse dynamics. Based on this, an Edgeworth expansion\footnote{The introduction of the Edgeworth expansion 
in cosmology context actually  dates back to the mid-nineties with \cite{Juszkiewicz95}
where it is introduced as a function basis decomposition as in the original references and \cite{Bernardeau95}
where it is based on an expansion around the Gaussian solution of the Laplace inverse
transformation so starting with our equation~\eqref{PDFfromphi}.} was performed by expanding the normalized saddle point PDF around a Gaussian in \cite{Gaztanaga00}. More recently, the leading order of the cumulant generating function has been obtained from a diagrammatic approach to the hydrodynamic equations in \cite{Bernardeau92} that was shown to lead to spherical collapse dynamics. Subsequently, in \cite{Valageas2002a}, spherical collapse has been shown to give the leading contribution to the cumulant generating function by means of a steepest descent method (sometimes also referred to as saddle point approximation) providing a non-perturbative argument that goes beyond the diagrammatic approach initially employed. In the following, we will adopt the approach to rely on spherical collapse dynamics to obtain the decay-rate function.}
%


When the saddle point approximation is applicable, that is when the decay-rate function is convex ($\Psi''_R[\hrho]>0$), it provides a very good approximation to the exact numerical integration \citep{Bernardeau14}. However, as mentioned before, it has been shown in \cite{1994A&A...291..697B}  that typically there is a critical value for $\hrho_c$ at finite distance where 
$\Psi_R''[\hrho_c]=0$ above which the Legendre transform of $\Psi_R$ is not defined,  which prevents the practical  use of equation~(\ref{PDFfromPsi}). Although there exist
alternative forms to the PDF based on the behaviour of the cumulant generating function near its critical point, they are only accurate in the very high density regime and
do not encompass the intermediate region around $\hrho\approx \hrho_c$. The central point we make in this paper is that this difficulty can be alleviated with an adequate change of variable $\hat\rho\rightarrow\mu$ such as the log of the density, $\mu = \log\hat\rho$.
The  construction of the density PDF is then obtained with the following steps:
\begin{subequations}
\label{Prhofrommu}
\begin{eqnarray}
\label{eq:Pmu}
\mP_{R,\mu}(\mu)\dd \mu&=&\int_{-i {\infty}}^{+i \infty} \frac{\dd\lambda}{2\pi i} \exp[-\lambda\mu+\phi_{R,\mu}(\lambda)]\,,\\
\mP_R(\hrho)\dd\hrho&=&\mP_{R,\mu}(\log(\hrho))\frac{\dd\hat\rho}{\hat\rho}\,,
\label{eq:Prhofrommu}
\end{eqnarray}
\end{subequations}
with the further simplification brought by the saddle-point expression of the density PDF in equation~\eqref{Prhofrommu},
 which eventually leads to
\begin{equation}
\label{PDFfromPsi2}
\mP_{R}(\hat\rho) = \sqrt{\frac{\Psi''_R[\hrho]+\Psi'_R[\hrho]/\hrho}{2\pi}} \exp\left(-\Psi_R[\hrho]\right)\,.
\end{equation}

It has to be noted that the two formulae, equations~\eqref{PDFfromPsi} and \eqref{PDFfromPsi2}, are based on two distinct assumptions when extrapolated to finite values of $\sigma$. Namely either
$\phi_{R,\hrho}(\lambda)=\varphi_{R,\hrho}(\lambda\sigma^{2})/\sigma^{2}$ or 
$\phi_{R,\mu}(\lambda)=\varphi_{R,\mu}(\lambda\sigma^{2})/\sigma^{2}$. 
We will see more precisely in the following that 
although the contraction principle ensures that the scaled cumulant generating functions are both independent 
of such assumptions -- their limit is left unchanged for $\sigma\to 0$ -- this is not the case when the variance is finite. 
This is why one variable turns out to be a better choice than the other in practice.

\begin{figure}
\includegraphics[width=8.5cm]{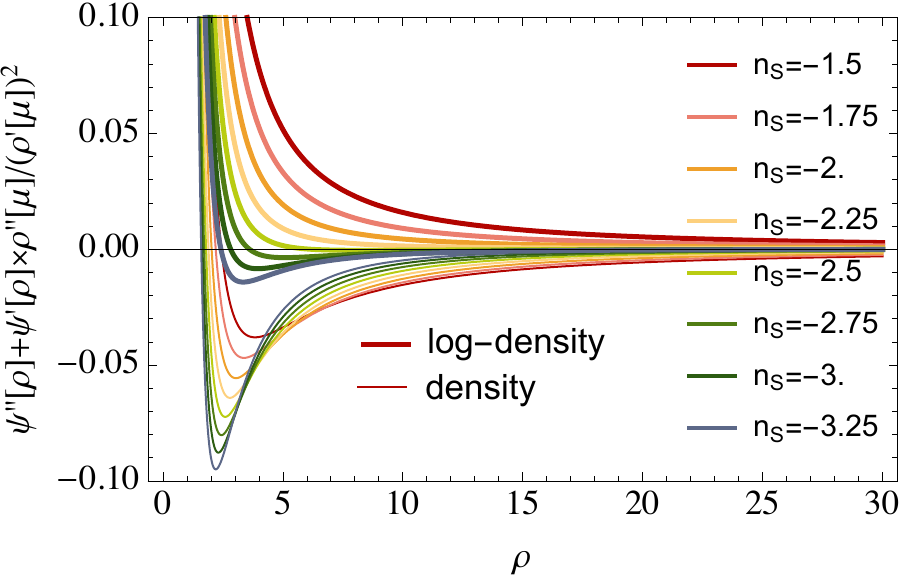}
\caption{Effect of a logarithmic density-transformation $\hrho(\mu)$ on the domain of definition of $\sqrt{\Psi''_\hrho}\ d\hrho$ set by the positivity condition $\Psi''_\hrho+ \Psi'_\hrho {\hrho''(\mu)}/{\hrho'(\mu)^2}>0$. Results for the density $\mu=\hrho$ ({\it thin lines}) and the log-density $\mu=\log\hrho$ ({\it thick lines}) are displayed for different initial spectral indices $n_s=-1.25,-1.5,\cdots,-3.25$ ({\it colored as indicated in the legend}). This comparison shows that the log-transform is able to avoid the criticality of the decay-rate function $\Psi_R$ for all densities over a wide range of indices typically $n_s\geq-2.4$. On the contrary, a critical point is met for all indices when the variable is the density field itself.}   
   \label{fig:PsiTrafo}
\end{figure}

Let us first examine the critical behaviour of $\Psi_{R}[\hat\rho]$ and $\Psi_{R}[\mu]$, respectively. The starting point is the quadratic form \eqref{Psiquad} for $\Psi_R(\tau)$ converted to a function of the final density $\hrho$ by inverting the spherical collapse relation to obtain $\tau(\hrho)$. For an EdS universe, the spherical collapse dynamics can be approximated as
\begin{align}
\hat\rho(\tau)\simeq \frac{1}{(1 - \tau/\nu)^\nu} \ , \ \nu = \frac{21}{13}\,,
\end{align}
\mod{which is known to reproduce well the spherical collapse in an EdS Universe \citep[described for example in ][]{Fosalba98} for the range of densities of interest as has been shown in \citep{Bernardeau92}.}
For simplicity, we now assume for the variance a power law initial power spectrum with index $n_s\approx-1.5$
\begin{align}
\label{sigsimple}
\sigma^2(R) = \sigma^2(R_p) \left(R/R_p\right)^{-(n_s+3)} \,,
\end{align}
where $R_p$ is a pivot scale. In Section~\ref{sec:PDFs} this simplifying assumption is amended to account for a running of the spectral index. In that case, the variance is approximated by
\begin{equation}
\sigma^{2}(R)=\frac{2\sigma^2(R_{p})}{(R/R_p)^{n_1+3}+(R/R_p)^{n_2+3}}\,,
\end{equation}
where $n_{1}$ and $n_{2}$ are chosen to reproduce the linear theory index $n(R)=-3-{\dd\log(\sigma(R))}/{\dd\log R}$  and running, $\alpha(R)={\dd\log(n(R))}/{\dd\log R}$ at the pivot scale $R_{p}$. For a generalization to arbitrary initial power spectra see Section~\ref{sec:outlook}.

The functions $\Psi_R''[\hat\rho]$ and $\Psi_R''[\mu]/{\hat\rho}^2 =\Psi_R''[\hat\rho] + \Psi_R'[\hat\rho] /{\hat\rho}$ entering the square root in respectively equation~\eqref{PDFfromPsi} and equation~\eqref{PDFfromPsi2} are shown in Fig.~\ref{fig:PsiTrafo} for various values of the power law index $n_s$. It can easily be checked that for most spectral indices of interest we always have $\Psi_R''[\mu]>0$.
Conversely, we recover the existence of a critical value $\hrho_c\simeq 2.36$ pointed out in \cite{Bernardeau14} for $\Psi_R[\hat\rho]$. 
One can see that the mapping $\hat\rho = \exp \mu$ avoids the criticality for all relevant densities and power spectrum indices $n_s\geq-2.4$.\footnote{For smaller indices $n_s<-2.4$ it is  possible to iterate the logarithmic mapping to prevent $\Psi_R$ from becoming critical.}

\section{Cumulant generating functions}
\label{sec:cumulants}

As stressed in the introduction, the application of the LDP gives access to the SCGF for the cumulants defined in  equation~(\ref{def:SCGF}) for the variable of interest. 
This quantity is at the heart of our constructions. It serves in particular as 
a model for the actual cumulant generating function -- which is an observable on itself -- as in equation~(\ref{hyp:CGF}). Such a function can be measured, or can be used to build the density PDFs 
as shown in the previous Section. 

\subsection{Scaled cumulant generating functions}

 Let us re-express the SCGF in terms of the field cumulants,
\begin{equation}
\label{exp:SCGF}
\varphi_{\hx}(\lambda)= \lim_{\langle\hx^{2}\rangle\to0}\ 
\sum_{p=1}^{\infty}\ \frac{\langle \hx^{p}\rangle_{c}}{\langle \hx^{2}\rangle_c^{p-1}}\,\frac{\lambda^{p}}{p!}\,,
\end{equation}
for a given variable $\hx$. It involves naturally the reduced cumulants $S_{p}[\hx]$ defined as
\begin{align}
S_p[\hx]= \frac{\langle\hx^p\rangle_c}{\langle\hx^2\rangle_c^{p-1}} \quad \forall p\geq 2 \,, 
\end{align}
but evaluated  
in their  low-variance limit.
Equation~(\ref{hyp:CGF}) contains however non-trivial physical assumptions.
From a PT point of view, and for Gaussian initial conditions as assumed here, the leading low-variance limit of $S_{p}[\hx]$
are their so-called tree order expression\footnote{This comes from their diagrammatic representations that all reduces to trees, see \cite{2002PhR...367....1B} for details.},
\begin{align}
 \lim_{\langle\hx^{2}\rangle\to0}\ S_p[\hx]= S_{p}^{\tree}[\hx].
\end{align}
The strength of the LDP applied in this context  is to provide means to compute {\sl all} the tree order expression of the reduced cumulant at once, for any variable such as $\hrho$ or $\mu$. Then the form (\ref{hyp:CGF}) relies on the hypothesis that either $S_{p}[\hrho]$ or $S_{p}[\mu]$ are {\sl independent} of variance, depending on the chosen variable.
Finally, notice that, in the specific case of the variable\footnote{Note that, $\mu=\log\hrho$ is used as shorthand notation for $\hrho=\exp\mu/\left\langle \exp \mu \right\rangle$ with $\langle\mu\rangle=0$.}  $\mu=\log \hrho$, there exists a simple way to compute
the moments of $\hrho$ from the cumulant generating function of $\mu$ as
\begin{equation}
\label{momfromSCGF}
\langle\hat\rho^{p}\rangle=\frac{\langle\rho^{p}\rangle}{\langle\rho\rangle^{p}}=\frac{\langle e^{p\mu}\rangle}{\langle e^{\mu}\rangle^p}=\exp[\phi_{\mu}(p)-p \,\phi_{\mu}(1)].
\end{equation}
It is then easy to predict the moments of the density from the SCGF. In particular, this allows to adjust the variance for $\hrho$, once the variance for $\mu$ has been chosen because 
\begin{equation}
\sigma_\hrho^2=\frac{\langle\rho^{2}\rangle}{\langle\rho\rangle^{2}}-1=\exp\left[\frac{\varphi_{\mu}(2\sigma_\mu^2)-2\varphi_{\mu}(\sigma_\mu^2)}{\sigma_\mu^2}\right]-1\,.
\end{equation}

\subsection{Cumulants as observables}
\begin{figure}
\centering
\includegraphics[width=0.45\textwidth]{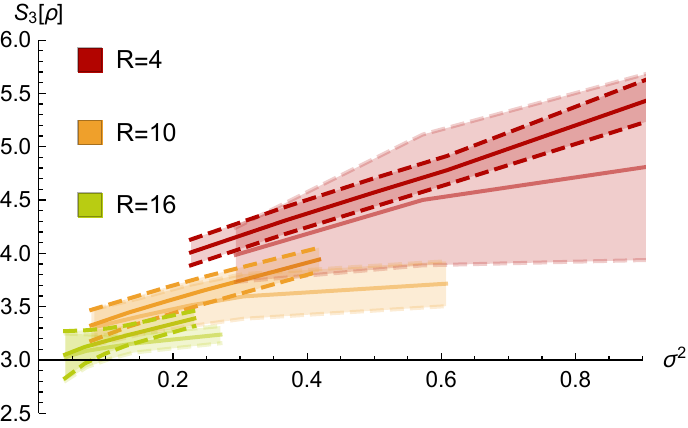}\\\vspace{0.2cm}
\includegraphics[width=0.45\textwidth]{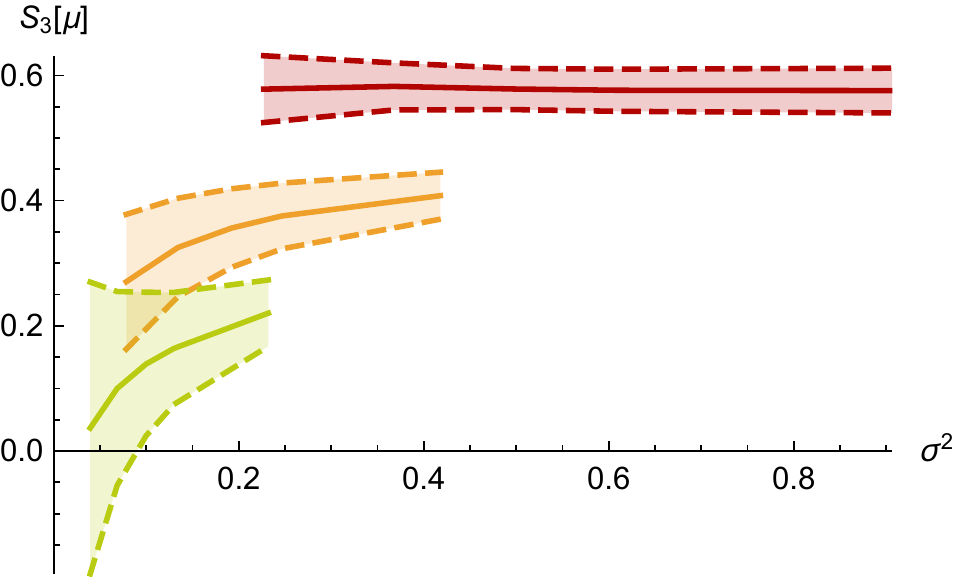}\\\vspace{0.2cm}
\includegraphics[width=0.45\textwidth]{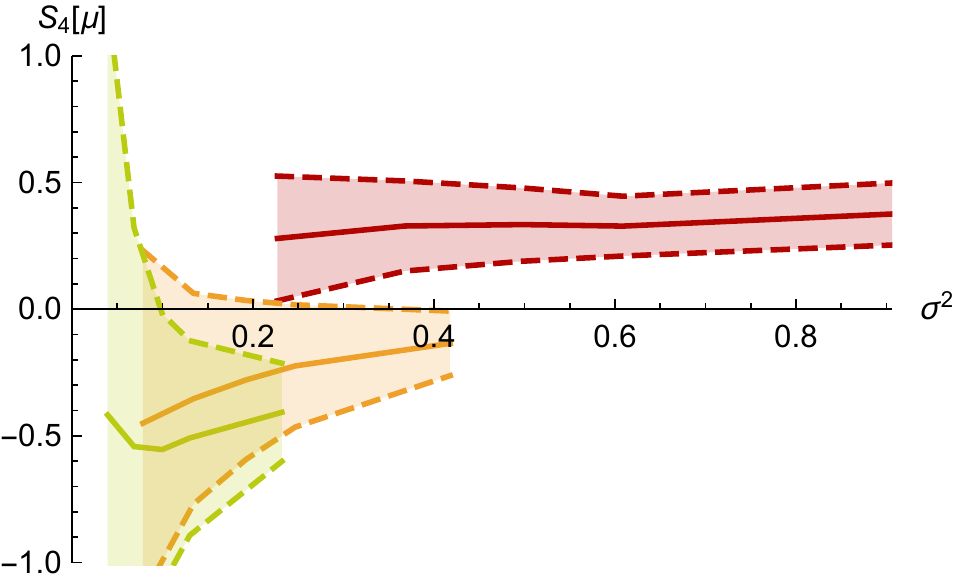}
  \caption{
Measured reduced cumulants with error bars (determined from 8 subsamples) versus the variance $\sigma^2_{\hrho}$ for $R$ in Mpc/$h$ as labeled. The measurements of $S_3[\hrho]$ {(\it top)}, 
$S_3[\mu]$ \textit{(middle)} and $S_4[\mu]$ \textit{(bottom panel)}, respectively, illustrate that the reduced cumulants of the log density $S_p[\mu]$ are almost constant while those of the density $S_p[\hrho]$  clearly change  with the variance. In the top panel, the direct measurement of $S_3[\hrho]$ ({\it lighter colour-shading}) is shown to be compatible with $S_3[\hrho]$ obtained from $S_{3/4}[\mu]$ according to formula \eqref{S3rhofrommu}.\vspace{-0.3cm}
}
\label{fig:measured-S3radii}
\end{figure}

From a theoretical point of view, the LDP does not give any indications about which physical assumption -- whether $S_{p}[\hrho]$ or $S_{p}[\mu]$ should be kept constant -- is better. PT calculations pushed to one-loop order could provide some indications, but no such results are known today. Hence, for now we must rely on results from numerical simulations. Those are described in Appendix~\ref{App:Simulation} \citep[see also][]{Bernardeau14}. In Fig.~\ref{fig:measured-S3radii}, we show the numerical variations of the reduced cumulants $S_3$ and $S_4$ for both $\hrho$ and $\mu=\log \hrho$ as a function of the variance for radii from $R=4$ to $16$ Mpc$/h$. We observe that the reduced cumulant for the log-density $S_p[\mu]$ is smaller than that of the density $S_p[\hrho]$, but also has a milder $\sigma$-dependence. This suggests that {extrapolating the zero variance result for the log-density to finite variances is more adequate than doing so for the density and also that} the cumulants of the log-density $S_p[\mu]$ can be better captured  by perturbation theory than those of the density $S_p[\hrho]$, as will be demonstrated in the following. 

\begin{figure}
\centering
\includegraphics[width=0.45\textwidth]{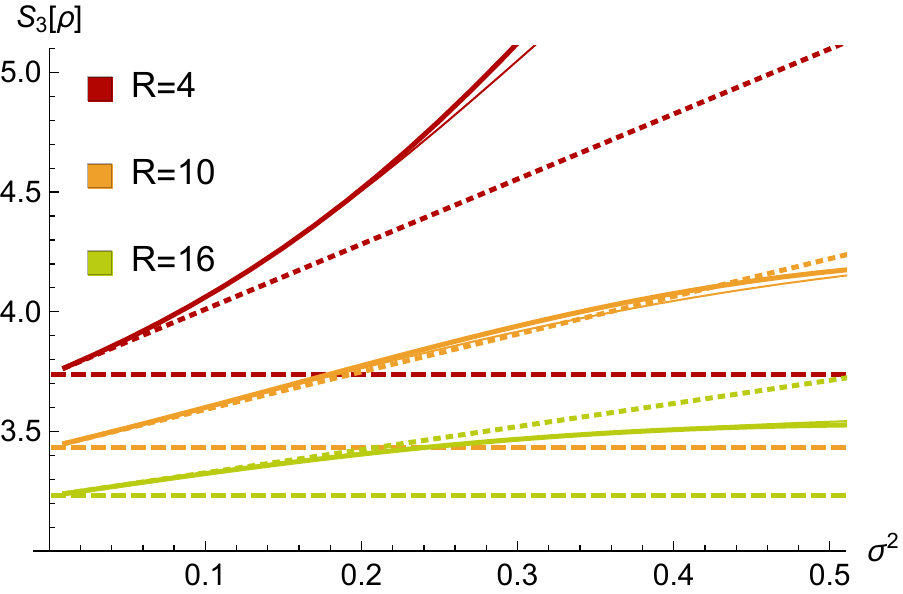}
\caption{\mod{Third reduced cumulant $S_3[\hrho]$ as a function of the variance $\sigma_{\hrho}^2$ obtained from the cumulant generating function using equation~\eqref{momfromSCGF} ({\it thick solid lines}) and from the saddle-point approximation of the PDF equation~\eqref{PDFfromPsi2} {\it (thin lines)} for three different radii $R$ in Mpc/$h$ in comparison to the tree-level PT prediction for $\hrho$ {\it (dashed)} and $\mu=\log\hrho$ translated to $\hrho$ according to equation~\eqref{rhofrommutreelevel}  {\it (dotted)}. Tree-level PT applied to $\mu=\log\hrho$ leads to a linear $\sigma^2$-dependence of  $S_3[\hrho]$ for small but finite variances.}}
\label{fig:S3sigmadep}
\end{figure}

Irrespectively of the choice we make, it is always possible to express the reduced cumulants of one variable
in terms of the other. The procedure is detailed in Appendix~\ref{App:CumMapping}. It relies solely on the mapping between the log-density $\mu$ and the normalized density $\hrho = \exp\mu/\langle\exp\mu\rangle$. We report here on some results showing the expression of the variance and first two nontrivial reduced cumulants 
\begin{align}
\label{sigmarhofrommu}
\sigma_{\hrho}^{2}=&\sigma _{\mu }^2+\left(S_3[\mu] +\frac{1}{2}\right) \sigma _{\mu }^4+\mathcal O(\sigma _{\mu }^6)\,,\\
 S_3[\hrho]=& S_3[\mu]+3 \label{S3rhofrommu}\\
&+\sigma^2_\mu\left(\frac{3}{2}S_4[\mu]+ 2S_3[\mu]-2(S_3[\mu])^2+1\right)+\mathcal O(\sigma _{\mu }^4)\,,\nonumber\\ 
S_4[\hrho]=&16+12 S_3[\mu]+S_4[\mu ] + \mathcal O(\sigma _{\mu }^2)\,. \label{S4rhofrommu}
\end{align}
In particular, equation~\eqref{S3rhofrommu} can be used to determine the third cumulant of the density $S_3[\hrho]$ from the measured cumulants of the log-density $S_{3/4}[\mu]$. This is illustrated in Fig.~\ref{fig:measured-S3radii} (top panel); it shows how well $S_3[\hrho]$ measured from the simulation can be recovered from the measured $S_{3/4}[\mu]$ and that the seemingly larger error bars on $S_{3/4}[\mu]$ for large radii indeed lead to comparable or even smaller error bars on $S_p[\hrho]$. 

Furthermore, assuming that the cumulants of the log-density $S_{p}[\mu]$ are equal to their tree order expression, allows to combine equations~\eqref{S3rhofrommu} and \eqref{S4rhofrommu} into
\begin{align}
\label{rhofrommutreelevel}
S_3[\hrho]=& S_3^{\tree}[\hrho]+\\ \nonumber
& \sigma^2_{\hrho} \left(\frac{3}{2} S_4^{\tree}[\hrho]-4 S_3^{\tree}[\hrho]-2(S_3^{\tree}[\hrho])^2+7\right)\,,
\end{align}
to first order in the variance. Assuming the $S_{p}[\mu]$ coefficients are constant then implies that those for the density are not.
The expected dependence based on tree-level perturbation theory is illustrated in Fig.~\ref{fig:S3sigmadep}. Here the expressions of the $S_{p}^{\tree}[\hrho]$ coefficients are computed for different radii $R$ and hence power law indices $n_s$ according to equations~(11)-(12) in \cite{Bernardeau14}. 
The result shows that tree-level perturbation theory in the log-density recovers the linear $\sigma^2$-dependence of the density result for the third reduced cumulant for small variances. This can be seen as hint that the log-density can prove useful for perturbation theory, at least as far as the highly symmetric setting of spherical collapse dynamics is concerned. Furthermore we show the results obtained from the Legendre transformation of the rate function equation~\eqref{phiLegendre} together with the cumulant relation equation~\eqref{momfromSCGF} and the PDF using the saddle-point approximation equation~\eqref{PDFfromPsi2}, respectively. The good agreement of the two methods point towards a wide applicability of the saddle-point PDF up to variances of $\sigma^2\simeq 0.5$ and provides an initial assessment independent of a numerical integration of equation~\eqref{Prhofrommu} that will be presented in Section~\ref{sec:onecell}.

\subsection{Cumulant generating functions as observables}

\begin{figure*}
\centering
\includegraphics[height=5.2cm]{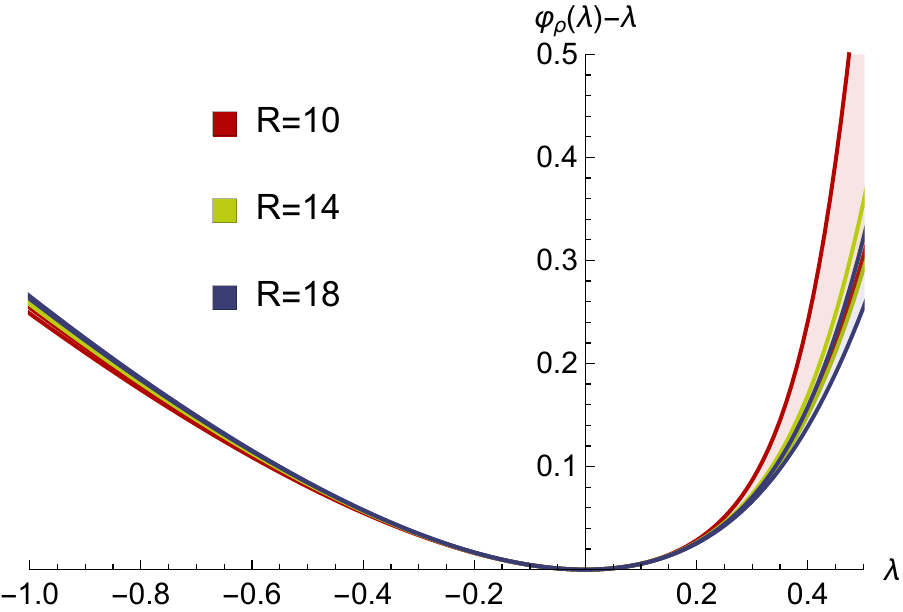} \quad \includegraphics[height=5.2cm]{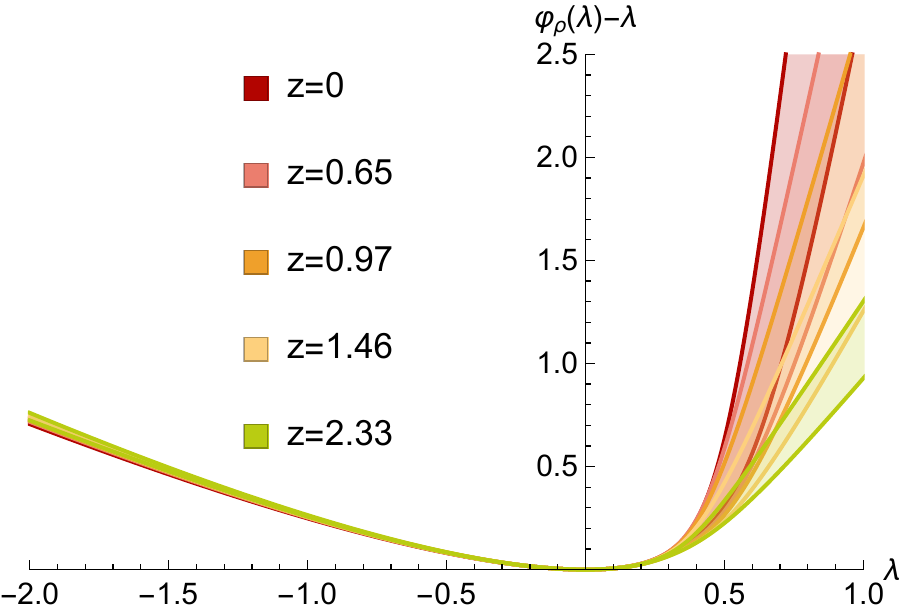}\\\vspace{0.2cm}
\includegraphics[height=5.2cm]{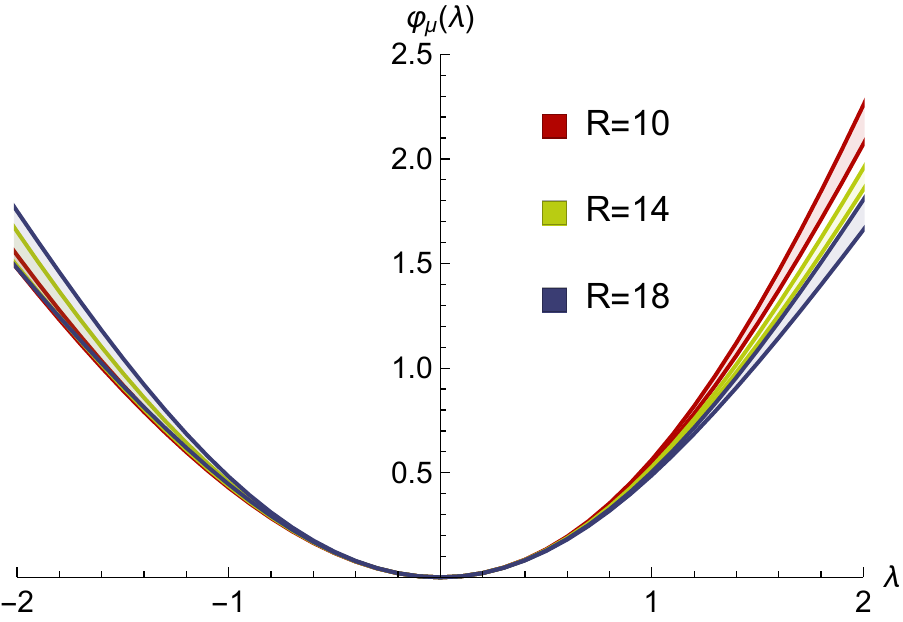} \quad \includegraphics[height=5.2cm]{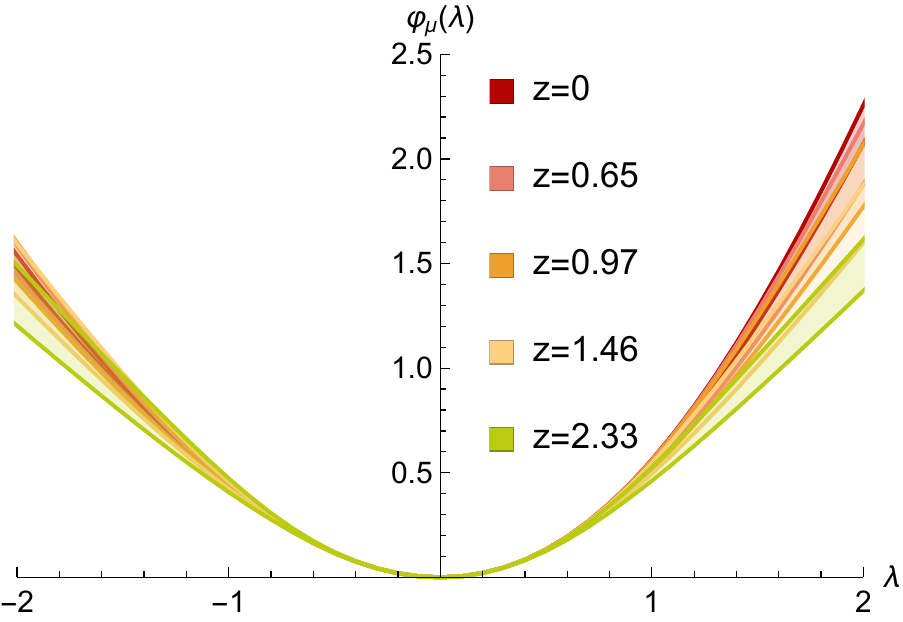}
  \caption{
\mod{Comparison of the measured SCGF $\varphi_\rho(\lambda)$ \textit{(top)} and $\varphi_\mu(\lambda)$ \textit{(bottom)} at redshift $z=0$ for radii $R$ (in Mpc/$h$) \textit{(left)} and for redshifts from $z=0$ to $z=2.33$ with the radius $R=10$ Mpc/$h$ \textit{(right)} as labeled. The considerably weaker dependence of $\varphi_\mu(\lambda)$ on the radius $R$ and the redshift $z$ (and hence the variance $\sigma^2$) shows visually that it is better justified to use the low variance limit result for $\varphi_\mu(\lambda)$ also for finite variances instead of doing so for $\varphi_\rho(\lambda)$. Note however that the same value $\lambda$ corresponds to different densities depending on whether one considers $\varphi_\rho(\lambda)$ or $\varphi_\mu(\lambda)$. This is illustrated in Fig.~\ref{fig:measured-rhomax-radii} using the saddle-point approximation.}}
 \label{fig:measured-phi}
\end{figure*}

\begin{figure}
\includegraphics[width=8cm]{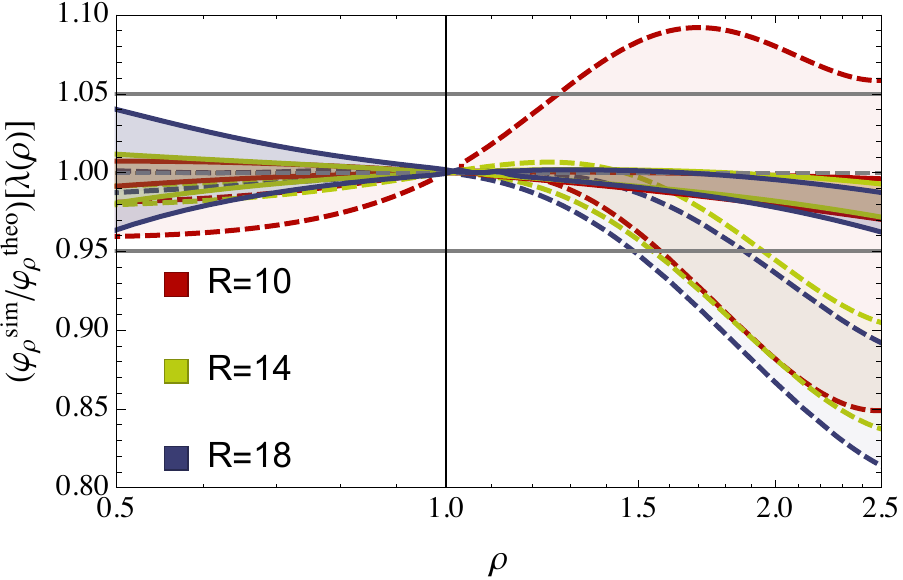}
\caption{The ratio of the measured SCGF and the prediction from the Legendre transform of the rate function \eqref{phiLegendre} plotted as a function of $\rho$ with the help of the saddle-point approximation $\lambda=\psi'$ for the density {\it (dashed lines)} and the log-density {\it (solid lines)}. Clearly, the residuals for the log-density are within 5\% for $\rho\in[0.5,2.5]$ while those for the density show besides criticality above $\rho_c\simeq 2.5$ also significant deviations for $\rho<\rho_c$.}
\label{fig:measured-vs-theory-phi}
\end{figure}

\begin{figure}
 \centering
\includegraphics[width=8cm]{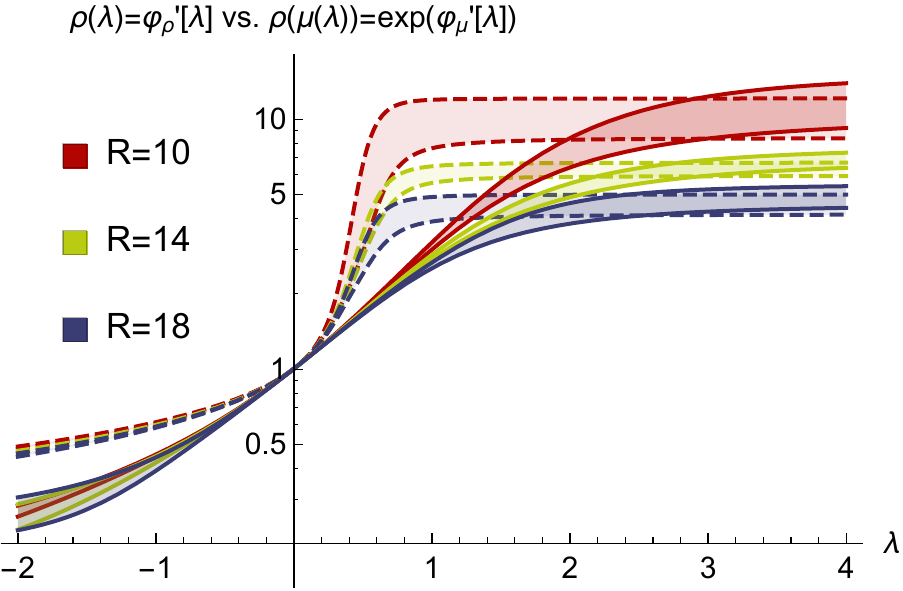}
  \caption{
Comparison of the saddle-point approximation for the SCGF showing $\rho=\varphi_\rho'(\lambda)$ \textit{(dashed lines)} and $\rho(\mu)=\exp \varphi_\mu'(\lambda)$ \textit{(solid lines)} at redshift $z=0$ for $R$ in Mpc/$h$ as labeled. The horizontal asymptotes of the derivative of the cumulant generating function $\varphi'$ at large $\lambda$ demonstrate that the Legendre transform will not be possible through that range of values. 
   \label{fig:measured-rhomax-radii}}
\end{figure}

Cumulant generating functions themselves are measurable in $N$-body simulations. In Fig.~\ref{fig:measured-phi} we show the scaled cumulant generating function $\varphi_R(\lambda)=\sigma^2\phi_R(\lambda/\sigma^2)$ for the log-density $\mu$ and the density $\hrho$ as measured from the $N$-body simulation for different radii and redshifts. For the SCGF of the density displayed in the upper panels the emergence of a critical point can be observed which makes the error bars large. Note that the dependence of the variance $\sigma^2$ on radius $R$ and redshift $z$ is such that decreasing $R$ and decreasing $z$ both increase the variance. Therefore a SCGF with a narrow band for different radii $R$ and redshifts $z$ signals robustness against increasing the variance from its zero limit upwards. As is evident from the four plots, the SCGF of the log-density reflects this property in contrast to the SCGF of the density which renders the latter unsuitable. 

Fig.~\ref{fig:measured-vs-theory-phi} compares the measurements to the theoretical result for the SCGF obtained from a Legendre transformation of the rate function according to equation~\eqref{phiLegendre}. The predictions are expected to be in good correspondence with the measurements if the SCGF is stable against changes in the variance. Again, the log-density clearly improves the range of applicability of the above-described  theoretical construction.

Fig.~\ref{fig:measured-rhomax-radii} shows the density which can be associated to the argument of the measured SCGF $\varphi(\lambda)$ from Fig.~ \ref{fig:measured-phi} by using the saddle-point approximation to obtain $\mu =\varphi_{\mu}'(\lambda)$ and from there $\hrho = \exp\mu$ or directly $\hrho = \varphi_\hrho'(\lambda)$. The saturation in the density as a function of $\lambda$ signals up to which maximum density the saddle-point approximation can be applied in principle. 

\begin{figure*}
\begin{minipage}[t][7.cm]{8.3cm}
\subfloat{\includegraphics[width=1.22\columnwidth]{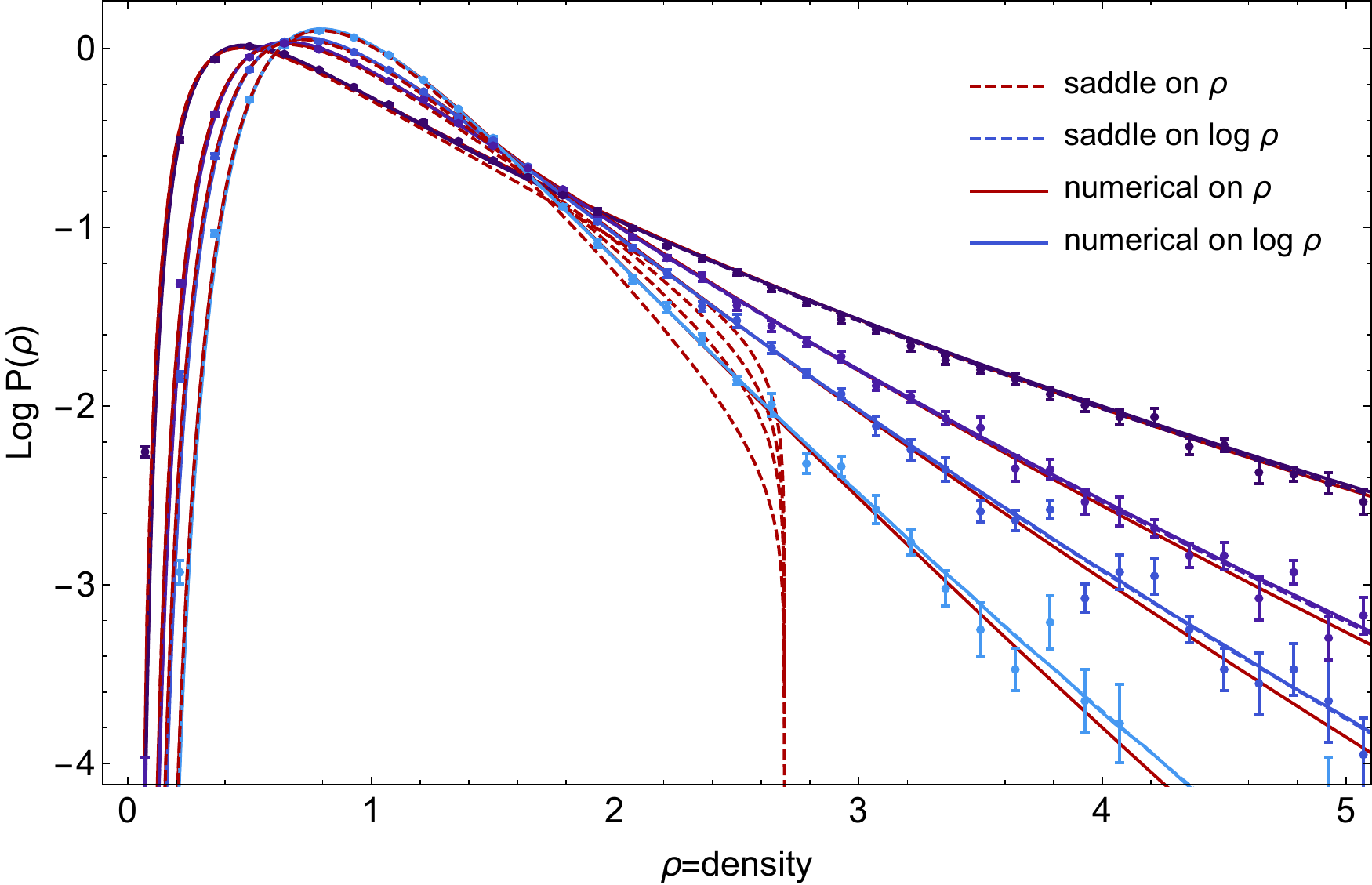}}
\end{minipage}
\hfill
\begin{minipage}[t][7.cm]{7.2cm}
\subfloat{\includegraphics[width=1.0\columnwidth]{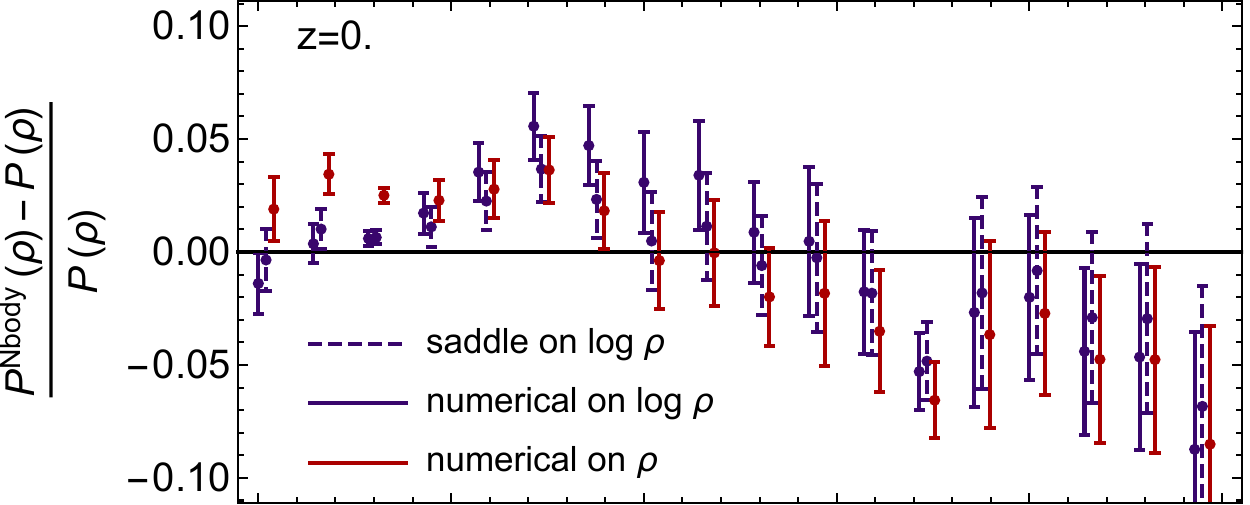}}\vspace{-0.44cm}
\subfloat{\includegraphics[width=1.0\columnwidth]{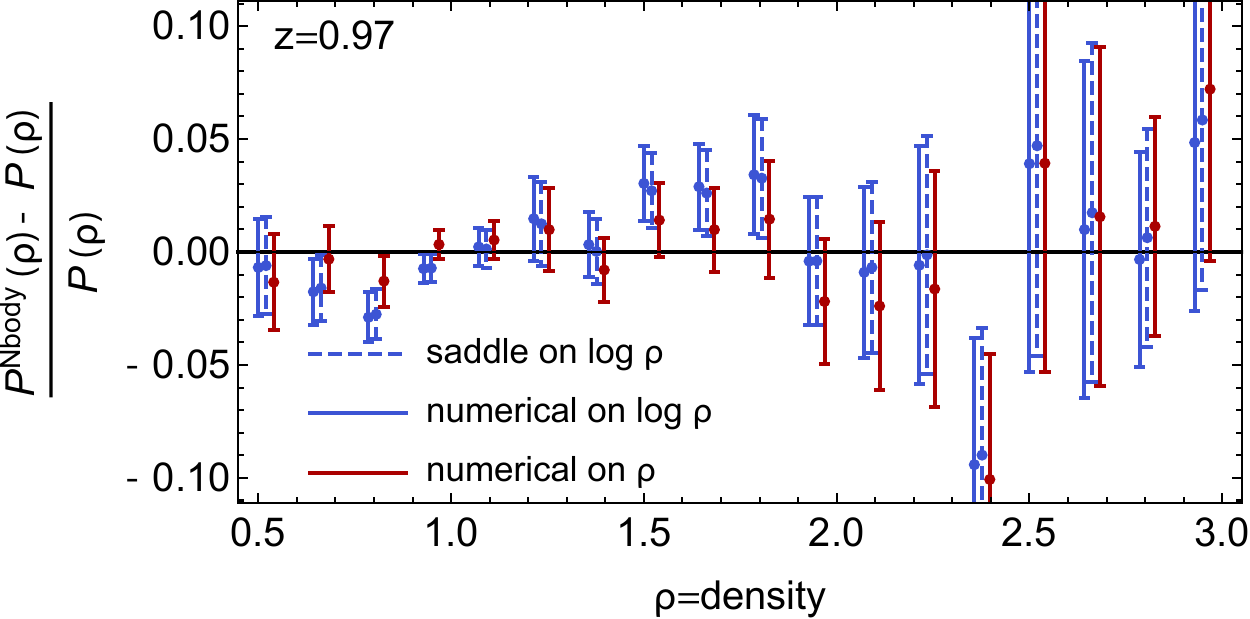}}
\end{minipage}
\caption{Left panel: PDF of $\rho$ measured {\it (points with error bars)} and predicted from a numerical integration in the complex plane for the density PDF following equation~\eqref{PDFfromphi} {\it(red solid lines)}, a numerical integration of the log-density PDF according to equations~\eqref{Prhofrommu} {(\it blue solid lines)}, a saddle-point approximation in the PDF of $\rho$ as written in equation~\eqref{PDFfromPsi} {\it(dashed red lines)} or a saddle-point approximation in the PDF of $\log \rho$ as mentioned in equation~\eqref{PDFfromPsi2} {\it (dashed blue lines)}. Four different redshifts are shown : $z=1.36$, 0.97, 0.65 and 0 {\it(from light to dark blue)}, for a filtering scale $R=10$ Mpc/$h$. \mod{The error bars represent the error on the mean as measured from 8 subsamples in our simulation.} The density PDF obtained from the PDF of $\log \rho$ has been rescaled in order to impose the normalisation, the mean and an effective variance has been used that allows us to recover the density variance measured in the simulation. Note that the solid and dashed blue lines are almost indistinguishable on this plot meaning that the saddle-point approximation gives a very accurate (and analytical!) fit to the  PDF when $\log \rho$ is taken as a variable.
Right panel: residuals at two different redshifts $z=$0 and 0.97 corresponding to $\sigma(R)=0.78$, and  0.48. The dashed blue and red  error bars have been shifted along the x-axis by respectively 0.02 and 0.04 for readability. Note that for densities below 0.5, the disagreement between prediction and measurement is larger and therefore not displayed here.}  
\label{fig:PDF}
\end{figure*}
\section{Constructions of the density PDFs}
\label{sec:PDFs}

We are now in position to build explicit density PDFs following the method sketched in Section~\ref{fb:theLDP}.

\subsection{One cell PDF}
\label{sec:onecell}

Fig.~\ref{fig:PDF} shows the density PDF obtained from the numerical integrations equations~\eqref{PDFfromphi} and \eqref{Prhofrommu} and the saddle-point approximations equations~\eqref{PDFfromPsi} and \eqref{PDFfromPsi2} in comparison to the measurement from the simulation including the corresponding residuals. It shows that if the log-density is used, the saddle-point approximation \eqref{PDFfromPsi2} (shown as blue dotted lines) becomes accurate over a wide range of densities $\rho\in [0.5,5]$ when compared to simulations, and coincides with the numerical integration for the log-density equation~\eqref{Prhofrommu} (shown as blue solid lines) for all densities. 

This success is the main result of the present work, as criticality is avoided for the relevant densities  when the log-density is considered.
It is to be contrasted with the case of the density where the saddle-point approximation equation~\eqref{PDFfromPsi} (shown as dashed red lines) is only applicable in the range $\rho\in [0.5,1.5]$ and a more demanding numerical integration in the complex plane must be implemented (shown as red solid lines) to evaluate the PDF using equation~\eqref{PDFfromphi} as shown in Fig.~2 in \cite{Bernardeau14}.

\subsubsection{Ensuring normalization}
Since the saddle-point method yields only an approximation to the exact PDF, the PDF obtained from equation\,\eqref{PDFfromPsi} has to be normalized $\hat\mP_{R}(\hat\rho) = {\mP_{R}(\hat\rho)}/{\displaystyle \int \mP_{R}(\hat\rho) \, d\hat\rho}$. Furthermore, when mapping a PDF for the log density $\mu$ with zero mean $\langle\mu\rangle=0$ to the one for a density using equation\,\eqref{PDFfromPsi2}, we have to consider the normalized density $\hat\rho=\rho/\langle\rho\rangle=\exp\mu/\langle\exp\mu\rangle$ in order to enforce $\langle\hat\rho\rangle=1$. Hence, the final PDF is obtained from equation\,\eqref{PDFfromPsi2} as
\begin{align}
\label{PDFfromPsinorm}
\hat\mP_{R}(\hat\rho)= \mP_{R}\left(\displaystyle \hat\rho \cdot \frac{ \displaystyle\int \hrho \, \mP_{R}(\hrho) \, d\hrho}{\displaystyle\int \mP_{R}(\hrho) \, d\hrho}\right)  \frac{\displaystyle\int \hrho \,\mP_{R}(\hrho) \, d\hrho}{\left(\displaystyle\int \mP_{R}(\hrho) \, d\hrho\right)^2}\,.
\end{align}

\subsubsection{Adjusting the variance}
The key parameter in the prediction of the PDF is the value of the variance at  the pivot scale. In practice, a possible strategy is to treat it  as a free parameter to be adjusted to the observations. But in principle, the variance $\sigma^{2}$ can also be predicted by linear theory 
\begin{equation}
\sigma^2(R) = \int\frac{\dd^{3} k}{(2\pi)^{3}}P^{\rm lin}(k)W_{\rm 3D}(k R)^2\,,
\label{eq:defSigma2}
\end{equation}
 where  $W_{\rm 3D}(k)$ is the shape of the top-hat window function in Fourier space,
 \begin{equation}
W_{\rm 3D}(k)=3\sqrt{\frac{\pi}{2}}\frac{J_{3/2}(k)}{k^{3/2}}\,,
\end{equation}
and $J_{3/2}(k)$ the Bessel function of the first kind of order $3/2$. 

In Table~\ref{tab:sigma} we show a comparison between the values for the variance depending on whether it is predicted by linear theory for a smoothing scale of $R=10$ Mpc/$h$ and a spectral index $n_s=-1.576$ or measured in the simulation for either $\mu$ or $\rho$. Note that, we also state the result from converting $\sigma_{\rho,\rm sim}^{2}$ to $\sigma_{\rho,\rm sim\rightarrow\mu}^{2}$ using equation~(\ref{sigmarhofrommu}) together with the tree-level perturbation theory result for the third reduced cumulant of the log-density $S_3^{\rm tree}[\mu]=S_3^{\rm tree}[\rho]-3$.
\begin{table}
\center
\begin{tabular}{l|c|c|c|c}
R=10  Mpc/$h$ &  $\sigma_{\rm lin}^2$ &$\sigma_{\mu,\rm sim}^2$ & $\sigma_{\rho,\rm sim}^2$ & $\sigma_{\rho,\rm sim\rightarrow\mu}^2$ \\\hline 
z=0.97 & 0.214 & 0.192 & 0.226 & 0.188 \\ 
z=0.65 & 0.286 & 0.247 & 0.305 & 0.242 \\
z=0.0 & 0.470  & 0.418 &  0.607 &  0.396
\end{tabular}
\caption{Comparison of variances between linear theory $\sigma_{\rm lin}^2$, measurements in the simulation for the log-density $\sigma_{\mu,\rm sim}^2$ and density $\sigma_{\rho,\rm sim}^2$ and the mapping $\sigma_{\rho,{\rm sim}\rightarrow\mu}^2$ from equation~\eqref{sigmarhofrommu}.}
\label{tab:sigma}
\end{table}
\begin{figure}
\centering
\includegraphics[width=7cm]{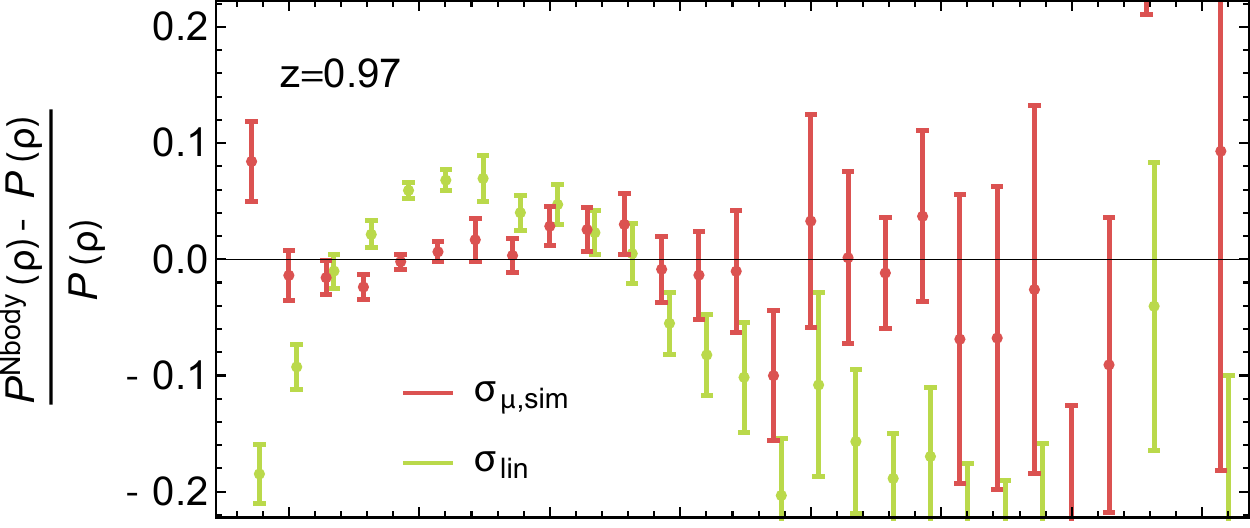}\\\vspace{-0.1cm}\ 
\includegraphics[width=7cm]{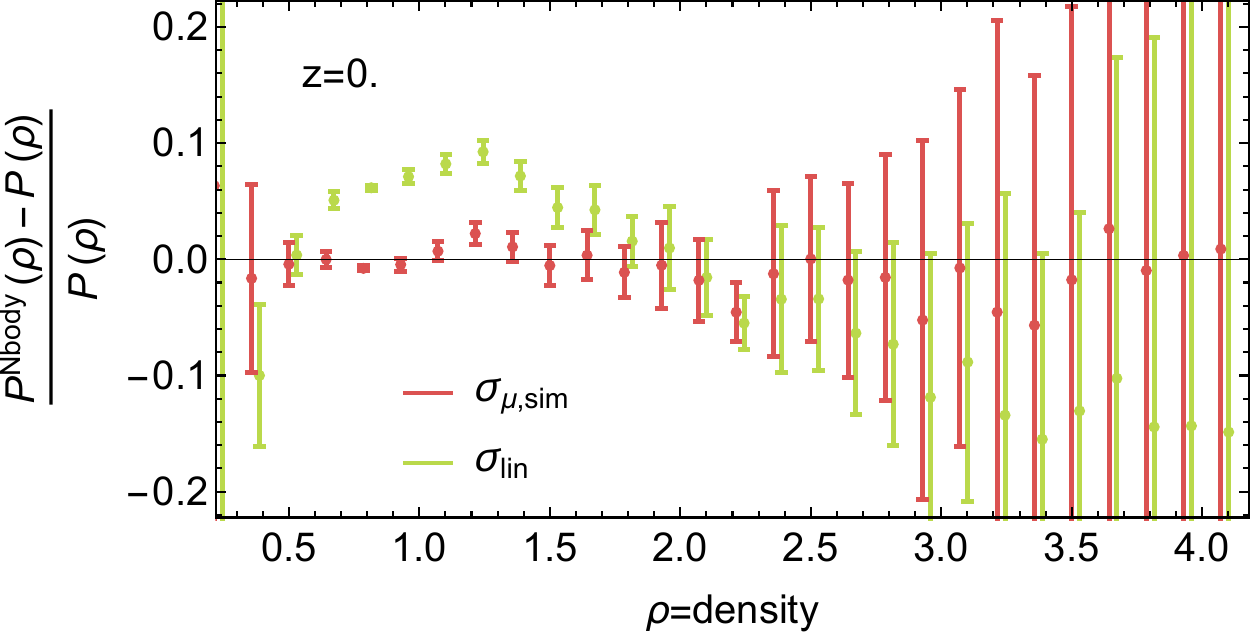}
\caption{Residuals for the measured PDF compared to two predictions for the log density using the measured $\sigma_{\mu,\rm sim}^2$ {(\it red)} or the linear result $\sigma_{\rm lin}^2$ {\it (green)} for two different redshifts as indicated. The value of $\sigma^2$ used here are extracted from table~\ref{tab:sigma}. Note that green error bars have been shifted along the x-axis by 0.03.}  
   \label{fig:PDFres-sigmavslin}
\end{figure}

\begin{figure*}
\includegraphics[width=0.9\columnwidth]{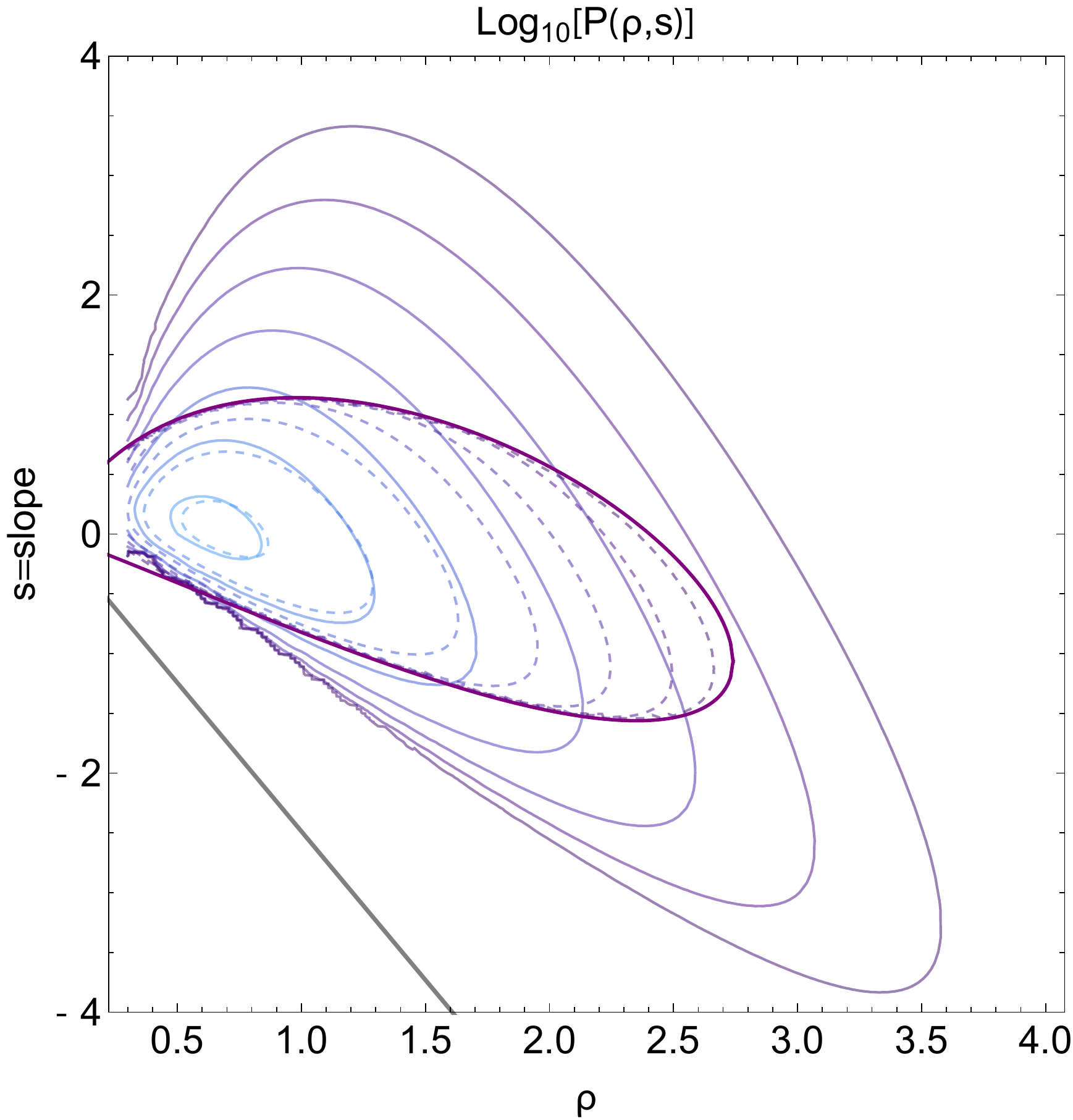}
\quad
\includegraphics[width=0.9\columnwidth]{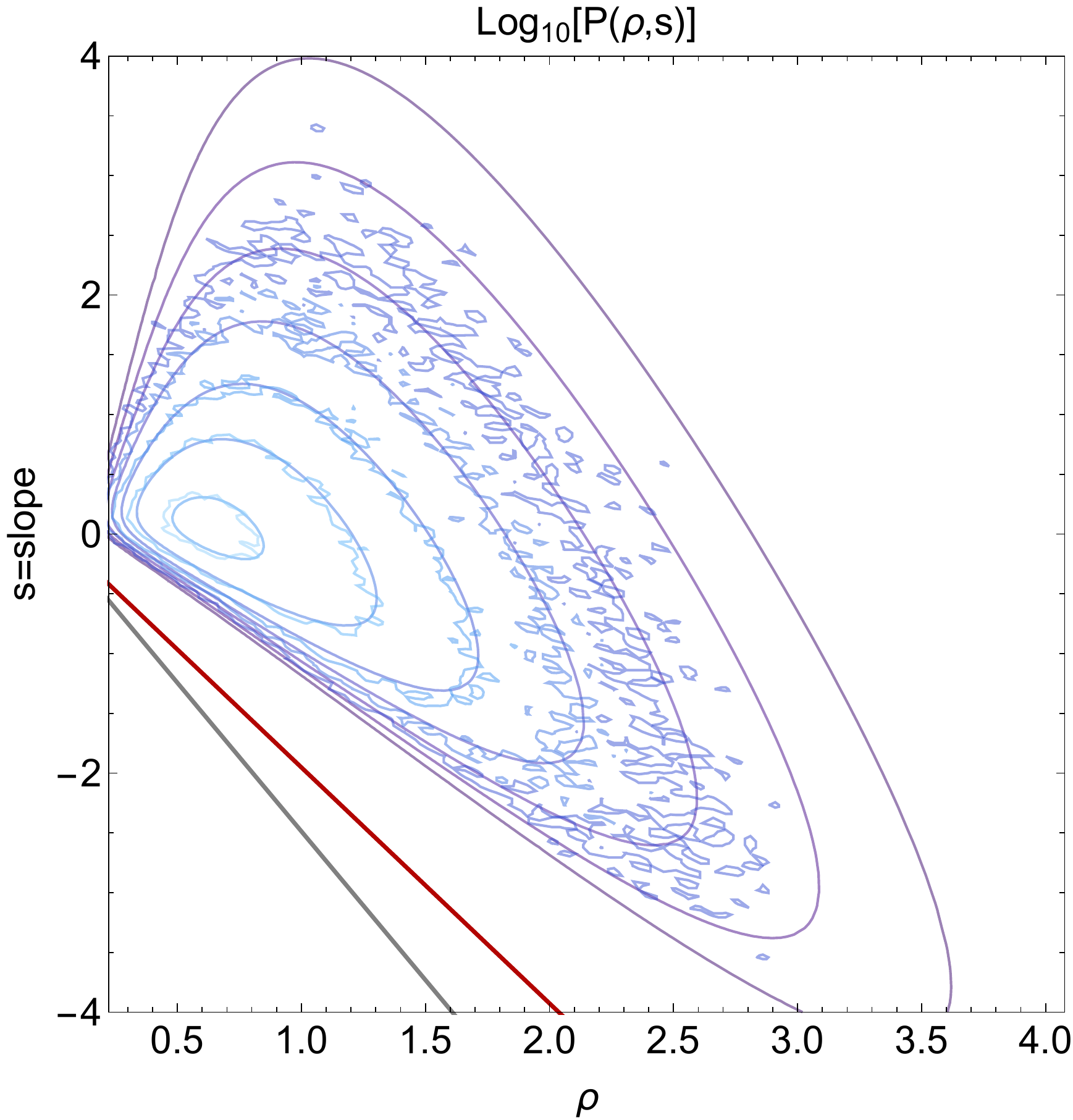}
   \caption{PDF of the inner density $\rho=\hrho_{1}$ and slope $s=(\hrho_{2}-\hrho_{1})R_{1}/(R_{2}-R_{1})$ within cells of radii $R_{1}=10$ and $R_{2}=11$Mpc$/h$ at redshift $z=0.97$. The {\it right panel} displays the log-mass saddle approximation given by equation~(\ref{eq:log-mass}) compared to the measured PDF while the {\it left panel} shows the numerical integration {\it (solid lines)} and density saddle approximation {\it(dashed lines)} of the joint PDF. Contours are displayed for ${\rm Log}\, \mP(\rho, s)=0,-1/2,-1,\dots,-3$. The grey solid line is the no-shell crossing limit $s>-10 \rho(1-r^{3})$, the red solid line is the critical line for the log-mass saddle approximation while the purple solid line is the critical line for the density saddle approximation.}
   \label{fig:PDF-2cell}
\end{figure*}

Finally, in Fig.~\ref{fig:PDFres-sigmavslin} we compare results obtained for the PDF for the log density using the saddle-point approximation depending on whether the linear variance is used for $\mu$ or the variance is measured from the simulation. Note that, while the linear prediction for the variance makes the model fully predictive without any free parameter, it systematically higher (around 10-15\% ) than the measured value, hence the prediction for the PDF is correspondingly not as accurate.

\subsubsection{Large density tail of the PDF}
Using the saddle-point approximation for the log density, equation~\eqref{PDFfromPsi2}, we can straightforwardly obtain the large density tails of the PDF as
\begin{align}
\label{LargeDensityTail}
\mathcal P_R(\hrho) \stackrel{\hrho\gg 1}{\longrightarrow} \frac{(n_s+3) \nu}{6 \sqrt{\pi \sigma_{\mu}^2(R)}} \exp\left[\!-\frac{\nu^2 (\hrho^{\frac{1}{\nu}}-1)^2\hrho^{\frac{n_s+3}{3}-\frac{2}{\nu}}}{2 \sigma_\mu^2(R)}\!\right]\hrho^{\frac{n_s-3}{6}}\,.
\end{align}
Equation~\eqref{LargeDensityTail} is surprisingly  simple and general w.r.t. the parameters of the theory, in contrast to the analytical asymptotic around the critical point $\hrho_c$ presented in \cite{Bernardeau14}, equation~(45). In particular, it shows explicitly how fitting the rare event tail of the PDF allows us to estimate $\nu$ and accordingly quantify possible modifications of gravity.

\subsection{The $2$-cell log density PDF saddle}
\label{sec:2cells}

Let us now explore the two-cell PDF  $\mP_{R_1,R_2}(\hat \rho_1,\hat \rho_2)$\ in the saddle approximation limit; this is a straightforward generalization of equation\,\eqref{PDFfromPsi} \citep[see][for the general expression of the 2-cells PDF]{bcp15}
\begin{equation}
 \mP_{R_1,R_2}(\hat \rho_1,\hat\rho_2)=
 \frac{ \exp\left[-\Psi_{R_1,R_2}(\hat \rho_1,\hat\rho_2)\right]}{2 \pi} {\sqrt{{\rm det}\!\left[\!\frac{\partial^{2}\Psi_{R_1,R_2}}{\partial\hat\rho_k \partial \hat \rho_l} \right]}} \,.  
\label{eq:joint2} 
\end{equation}

If the densities $\{(\hat \rho_1,\hat \rho_2)\}$ are used as variables, the issue of criticality for the Hessian 
\[{\rm det}\!\left[\!\frac{\partial^{2}\Psi_{R_1,R_2}}{\partial\hat\rho_k \partial \hat \rho_l}\right](\hat\rho_1,\hat\rho_2)_c=0\] 
becomes more severe compared the one-cell case where the saddle-point approximation broke down above a critical density. As demonstrated in the left panel of Fig.~\ref{fig:PDF-2cell} there is a roughly elliptical critical boundary $\{(\hat\rho_1,\hat\rho_2)_c\}$ beyond which the saddle-point method breaks down. Since the slope, given by the difference between the central and the overall density, is much more restricted, this suggests to apply the logarithmic transform not to the densities individually but to their difference and sum. 

A suitable and physically motivated choice for the difference is a mass-weighted one which ensures a well-behaved logarithm as long as the no-shell crossing condition $R_2^{3}\hrho_{2}- R_1^3\hrho_{1}>0$ is satisfied. This suggest to perform the following logarithmic transform of the sum and difference of mass
\begin{subequations}
 \label{log2cell}
 \begin{align}
 \mu_{1}&= \log  \left(r^{3}\hrho_{2}+\hrho_{1}\right)  \,,\\
   \mu_{2}&= \log \left(r^{3}\hrho_{2}-\hrho_{1}\right)\,,
 \end{align}
 \end{subequations}
 where the relative shell thickness is $r=R_{2}/R_{1}$ and the no-shell crossing condition enforces $\mu_{2}$ to be real. The PDF $\mP(\hrho_{1},\hrho_{2})$, or equivalently $\mP
 (\rho,s)$ the PDF of the inner density $\rho=\hrho_{1}$ and slope $s=(\hrho_{2}-\hrho_{1})/(r-1)$, can then be approximated via a saddle-point approximation by
 \begin{align}
 \label{eq:log-mass}
 \mP_{R_1,R_2}(\hat\rho_{1},\hat\rho_{2})&= \frac{ \exp\left[-\Psi_{R_1,R_2}(\hrho_{1},\hrho_{2})\right]}{ 2 \pi} \\
&\quad \times \sqrt{\det\left[\frac{\partial^{2}\Psi_{R_1,R_2}}{\partial \mu_{i}\partial \mu_{j}}\right]}\left|\det\left[\frac{\partial\mu_{i}}{\partial \hrho_{j}}\right]\right| 
\,,\nonumber 
 \end{align}
 which can explicitly be rewritten as 
  \begin{equation*}
 \mP_{R_1,R_2}(\hat\rho_{1},\hat\rho_{2})=
\frac{ \exp\left[-\Psi_{R_1,R_2}(\hrho_{1},\hrho_{2})\right]}{ 2 \pi} \sqrt{ p_{R_1,R_2}(\hat\rho_{1},\hat\rho_{2})}\,,
 \end{equation*}
 with
 \begin{align}
p_{R_1,R_2}(\hat\rho_{1},\hat\rho_{2})=&\det\left[\frac{\partial^{2}\Psi_{R_1,R_2}}{\partial \mu_{i}\partial \mu_{j}}\right]\left(\det\left[\frac{\partial\mu_{i}}{\partial \hrho_{j}}\right]\right)^{2}\nonumber\\
=
 &
 \left(\frac 1 {2r^{3}} \Psi_{,22}+\Psi_{,12}+\frac {r^{3}} {2}\Psi_{,11}+\frac{\Psi_{,2}+r^{3}\Psi_{,1}}{r^{3}\hrho_{2}+\hrho_{1}}\right)\nonumber\\
 \times& \left(\frac 1 {2r^{3}} \Psi_{,22}-\Psi_{,12}+\frac {r^{3}} {2}\Psi_{,11}+\frac{\Psi_{,2}-r^{3}\Psi_{,1}}{r^{3}\hrho_{2}-\hrho_{1}}\right)\nonumber
\\
-&\left(\frac {\Psi_{,22}}{2 r^{3}}-\frac{r^{3}\Psi_{,11}}{2}\right)^{2}\,,
 \end{align}
with $\Psi_{,1}$ and $\Psi_{,2}$ denoting partial derivatives with regard to $\hrho_{1}$ and $\hrho_{2}$ respectively.

\begin{figure}
\hspace{0.3cm} \includegraphics[width=0.95\columnwidth]{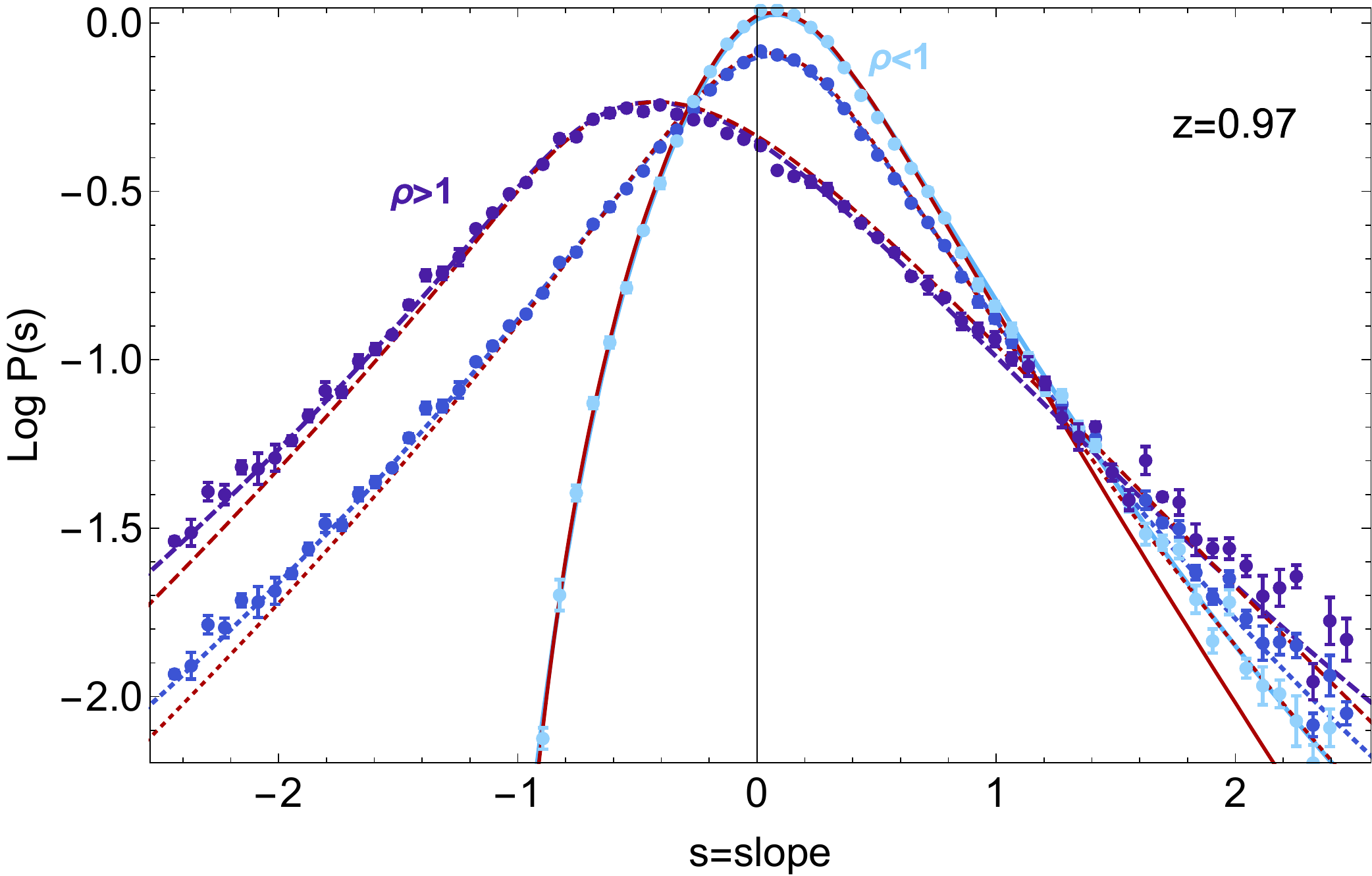}
\includegraphics[width=\columnwidth]{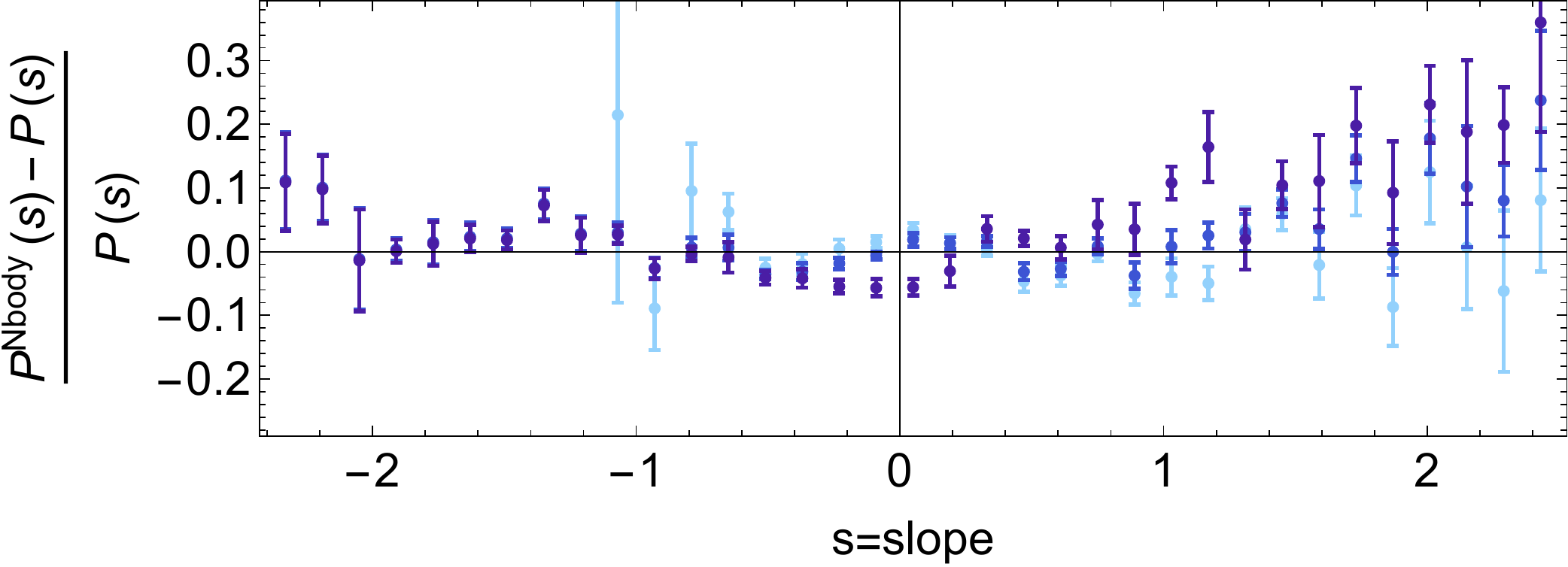}
   \caption{{\it Top panel:} PDF of the slope, of the slope when the inner density is below one and of the slope when the inner density is above one. Error bars represent the error on the mean as measured \mod{from subsamples} in our simulation, red lines represent the numerical integration while blue lines are the log-mass saddle approximation given by equation~(\ref{eq:log-mass}). 
   The agreement is very good for the whole range of density and slope probed by the simulation.
  {\it  Bottom panel:} residuals of measured slope PDFs compared to the log-mass saddle approximation corresponding to the blue lines in the top panel.
   }
   \label{fig:Ps-2cell}
\end{figure}

 This change of variables allows to get analytical approximations valid for a wide range of densities and variances. In particular, for a variance $\sigma=0.48$, the right panel of Fig.~\ref{fig:PDF-2cell} shows that the critical line (in red) only excludes a marginal fraction of the $\rho-s$ plane (between the red and the grey lines) which has very little weight ($\mP\approx 0$ in those regions).
The full joint PDF of concentric densities and slopes computed from equation~(\ref{eq:log-mass}) is also shown in  Fig.~\ref{fig:PDF-2cell} while Fig.~\ref{fig:Ps-2cell} uses the two-dimensional knowledge of the PDF to predict the PDF of the slope in subregions (under-dense, over-dense or unconstrained inner cells). The agreement between measurements and the analytical predictions given by equation~(\ref{eq:log-mass}) is remarkably good, even better than the numerical integration of \cite{bcp15}, which probably suffers from numerical inaccuracies in the rare event tails of the distribution.
The success of this analytical approach is to be contrasted with the severely limited range of validity of the saddle-point approximation of the density PDF illustrated in the left panel of Fig.~\ref{fig:PDF-2cell}.
  
\section{Conclusion}
\label{sec:conclusion}
\subsection{Summary}
The large deviation principle allows us to make simple and accurate predictions for the cumulants of the distribution of the density within concentric shells based on spherical collapse dynamics. Using the log density considerably extends the regime where the PDF derived from the saddle-point approximation matches the exact PDF because it remedies the problem of criticality reported before.

In particular, the simple analytic model is shown to be able to match the PDF of the density for all densities when compared with a numerical integration in the complex plane. The result for the log-density can be easily linked to the PDF of the density as a one-line approximation, equation~\eqref{PDFfromPsi2}
 and also yields excellent results over a large range of density values when compared to measurements from $N$-body simulations as illustrated in Fig.~\ref{fig:PDF}. In particular, this expression gives immediate access to the rare event tail of the density PDF for large positive densities.
The two cells joint PDF of the log density was also presented in the saddle approximation limit. 
The mass weighted logarithmic mapping performed according to equation\,\eqref{log2cell} yields also an analytic PDF and works very well, making almost the entire space of density and slope $\{\rho,s\}$ accessible as shown in Fig.~\ref{fig:PDF-2cell}.

{The origin of the success of the log-density lies in the applicability of its saddle-point approximation and is supported by the quality of the Ansatz corresponding to equation~(\ref{hyp:CGF}), as the cumulants of the log transformed field depend more weakly on their (finite) variance, as illustrated in Fig.~\ref{fig:measured-S3radii}. This can also render tree-level perturbation theory in the log density more successful in predicting reduced cumulants of the density.}

\subsection{Perspectives for Dark Energy}
\label{sec:outlook}

Statistics for densities in concentric shells will prove very useful in upcoming surveys as they allow us to study the clustering of peaks (or voids) 
in  the mildly non linear regime ($\sigma\sim1$) and
serve as a statistical indicator to test gravity and dark energy models and/or probe key cosmological parameters in carefully chosen subsets of surveys.  

A clear asset of the analytical saddle approximation is that it provides means of simply 
probing the variation of counts in cells for {\sl arbitrary} initial power spectra and spherical collapse models, which is clearly of interest in 
the context of dark energy/modified gravity investigations. In particular, it has to be noted that unlike the numerical integration in the complex plane given by equation~\eqref{PDFfromphi}, the saddle-point method equations~(\ref{PDFfromPsi2}) in the one-cell case and (\ref{eq:log-mass}) in the two-cell configuration do not require an analytical linear power spectrum. In particular, $\Lambda$-CDM-like power spectra can be used in this context. 
Recall that the  knowledge of the linear power spectrum determines the values of the cross-correlation matrix elements, $\Sigma_{ij}$, that are explicitly given by
\begin{equation}
\Sigma_{ij}(R_{i},R_{j})=\int\frac{\dd^{3} k}{(2\pi)^{3}}P^{\rm lin}(k)W_{\rm 3D}(k R_{i})W_{\rm 3D}(k R_{j})\,.
\label{eq:defSigmaij}
\end{equation}
In particular, we have
\begin{equation}
\frac{\partial^{}  \Sigma_{ij}}{\partial\hat\rho_k}= \frac{1}{3 \rho_k^{2/3}}\left({\delta_i^k}+
{\delta_j^k}
\right) \frac{\partial^{}  \Sigma_{ij}}{\partial R_k}\,,
\end{equation}
and 
\begin{align}
\frac{\partial^{2}  \Sigma_{ij}}{\partial\hat\rho_k\hrho_l}=& \frac{\left({\delta_i^k}+
{\delta_j^k}
\right)\left({\delta_i^l}+
{\delta_j^l}
\right)}{9 \rho_k^{2/3} \rho_l^{2/3}} \frac{\partial^{2}  \Sigma_{ij}}{\partial R_k\partial R_l}
 \nonumber \\
 &-
 \frac{2 \delta_l^k}{9 \rho_k^{5/3}}\left({\delta_i^k}+
{\delta_j^k}
\right) \frac{\partial^{}  \Sigma_{ij}}{\partial R_k}\,.
\end{align}
Now given equation~(\ref{eq:defSigmaij}), we get
\begin{align}
\frac{\partial  \Sigma_{ij}}{\partial R_l}=&\!\!\int\!\!\frac{\dd^{3} k}{(2\pi)^{3}}P^{\rm lin}(k) k W'_{\rm 3D}(k R_l) W_{\rm 3D}(k R_{(i}) \delta^l_{j)} \,,
\nonumber \\
\frac{\partial^2  \Sigma_{ij}}{\partial R_l\partial R_k}=&\!\!\int\!\!\frac{\dd^{3} k}{(2\pi)^{3}}P^{\rm lin}(k) k^2 \left[ W''_{\rm 3D}(k R_{k}) W_{\rm 3D}(k R_{(i}) \delta^k_{j)} \delta^{k}_{l}   \right.
\nonumber\\
&\qquad\qquad\left. +W'_{\rm 3D}(k R_{k})W'_{\rm 3D}(k R_{l})\delta^k_{(i}\delta^l_{j)}\right]
\,, \label{eq:dsigdR}
\end{align}
where $A_{(k}B_{j)}= A_kB_j+ A_jB_k$.
Hence equations~(\ref{eq:log-mass}) and (\ref{eq:dsigdR}) yield for instance an explicit expression for the PDF of the density in $2$ cells
in terms of an underlying arbitrary linear power spectrum $P^{\rm lin}(k) $.
 In practice, it will allow us to quantify very accurately the cosmological information contained in concentric cell observables. However, this is beyond the scope of this paper and therefore left for future works. \mod{For a foretaste of concrete applications we refer to \cite{Codis16b} where a proof of principle for constraining dark energy is provided and the accompanying code LSSFAST\footnote{LSSFAST: \url{http://cita.utoronto.ca/~codis/LSSFast.html}} to compute density PDFs for arbitrary initial power spectra.}

This formalism together with the log transformations could also be directly extended to the joint log density-velocity field divergence statistics, or more generally to any observable which can be expressed, not necessarily linearly, in terms of the densities in concentric cells. \mod{In order to make the present formalism ready-to-use for counts-in-cell determined from real datasets, galaxy bias and redshift space distortions have to be taken into account. The complications arising from redshift space distortions could possibly be circumvented by applying the formalism to projected cells with cylindrical symmetry corresponding to photometric redshift surveys. For first results regarding the effects of spatial correlations between cells at finite distance, see \cite{Codis16a}. }
\vspace{0.5cm}

{\bf Acknowledgements:}  
 This work is partially supported by the grants ANR-12-BS05-0002 and  ANR-13-BS05-0005 of the French {\sl Agence Nationale de la Recherche}. CU is supported by the Delta-ITP consortium, a program of the Netherlands organization for scientific research (NWO) that is funded by the Dutch Ministry of Education, Culture and Science (OCW). CU thanks IAP for hospitality when this project was initiated and the Balzan Centre for Cosmological Studies for financial support during the visit. The simulations were run on the {\tt Horizon} cluster. We acknowledge support from S.~Rouberol for running the cluster for us.
 
\bibliographystyle{mn2e}
\bibliography{LSStructure}

\appendix
\section{Path integral derivation}
\label{App:Intuitive}
Let us present an alternative derivation of the theoretical construction that yields the PDF $\mP_{R}(\hrho)$. In fact, the steepest decent method can be used to compute the cumulant generating function in terms of a Legendre transform of a function defined in terms of the initial density field.
Recall first that the statistical properties of $\hrho$ are fully encoded in its moment generating function
\begin{align}
\mgen_{R}(\lambda)&= \langle \exp(\lambda\hrho)\rangle 
=\int\dd\hrho \  \mP_{R} (\hrho)  \exp(\lambda\hrho)\,,
\label{eq:mgen1}
\end{align}
where $\mP_{R}(\hrho)$ is the PDF of having density $\hrho$ in ${\cal S}$. The moment generating function is related to the {\sl cumulant} generating function, $ \phi_{R}(\lambda)$, through 
\begin{equation}
\mgen_{R}(\lambda)= \exp\left[ \phi_{R}(\lambda)\right] \,.
\label{momcumrel}
\end{equation}
Now it is  always possible to re-express formally any ensemble average such as equation~\eqref{eq:mgen1}
 in terms of the statistical properties of the {\it initial} density field $\tau$, so that we can formally write
\begin{equation}
\hskip -0.1cm
\exp\left[ \phi_{R}(\lambda)\right]\!\!=\!\! \int\!\!\mD\tau\,\mP_R(\tau)
 \exp\big(\lambda\hrho(\tau)\big)\,. \hskip -0.3cm
\label{phiexp2}
\end{equation} 
As the present-time density $\hrho$ can arise from different initial contrasts $\tau$, the above-written integral should therefore  be understood a path integral over all the possible paths from initial conditions to present-time configuration, with measure $\mD\tau$ and known initial statistics $\mP(\tau)$. Let us assume here that the initial PDF is Gaussian so that,
\begin{equation}
 \mP_R(\tau)\dd\tau=
\frac{\exp\left[-\Psi_R(\tau)\right]}{\sqrt {2\pi \sigma^2(\tau)}}\dd\tau\,,
\label{mPexp1}
\end{equation}
with $\Psi_R$ then a quadratic form
\begin{align}
\label{Psiquada}
\Psi_R(\tau(\hrho)) = \frac{1}{2\sigma^2(\tau)} \tau(\hrho)^2\,.
\end{align}

In the  regime where the variance of the density field is small,
 equation~(\ref{phiexp2}) should be dominated by the path corresponding to the most likely configurations (or in the language of LPD the least unlikely of all 
 unlikely configurations).
 As the constraint is spherically symmetric, this most likely path should also satisfy  spherical symmetry. It is therefore bound to obey 
 the spherical collapse dynamics. Within this regime equation~(\ref{phiexp2}) becomes approximatively 
  \begin{equation}
  \hskip -0.1cm
\exp\left[ \phi_{R}(\lambda)\right] \simeq \int\!\! d\tau \,\mP_R(\tau)
 \exp\big(\lambda\zeta_{{\rm SC}}(\tau)\big),\hskip -0.45cm
\label{mPexp3}
\end{equation}
 where the most likely path, $\hrho=\zeta_{\rm SC}(\eta,\tau)$ is the one-to-one spherical collapse mapping between one final density at time $\eta$ and one initial density contrast as already described.
Putting equation~\eqref{mPexp1} into equation~\eqref{mPexp3},
the integration on the r.h.s. of equation~(\ref{mPexp3}) can now be carried by using 
a steepest descent method, approximating the integral as its most likely value,  where $\phi_{R}(\lambda) = \lambda\hrho(\tau)- \Psi_R(\tau) $ is 
stationary. 
If $\Psi(\hrho)$ is convex, then the Legendre transform is involutive, what implies that 
$\Psi_{R}(\hrho)$ is the \textsl{Legendre transform} of $\phi_{R}(\lambda)$,
 where $\rho$ is determined implicitly by the stationary condition
\begin{equation}
\Psi_{R}(\hrho)=\lambda\hrho-\phi_{R}(\lambda),\label{phifromLeg}\quad 
\lambda=\frac{\partial}{\partial \hrho} \Psi_{R}(\hrho)\,
\,.
\end{equation}
Equation~\eqref{phifromLeg} is of course fully consistent with the more general result derived from LDP.

\section{Simulation}
\label{App:Simulation}
\begin{figure}
 \center\includegraphics[width=0.90\columnwidth]{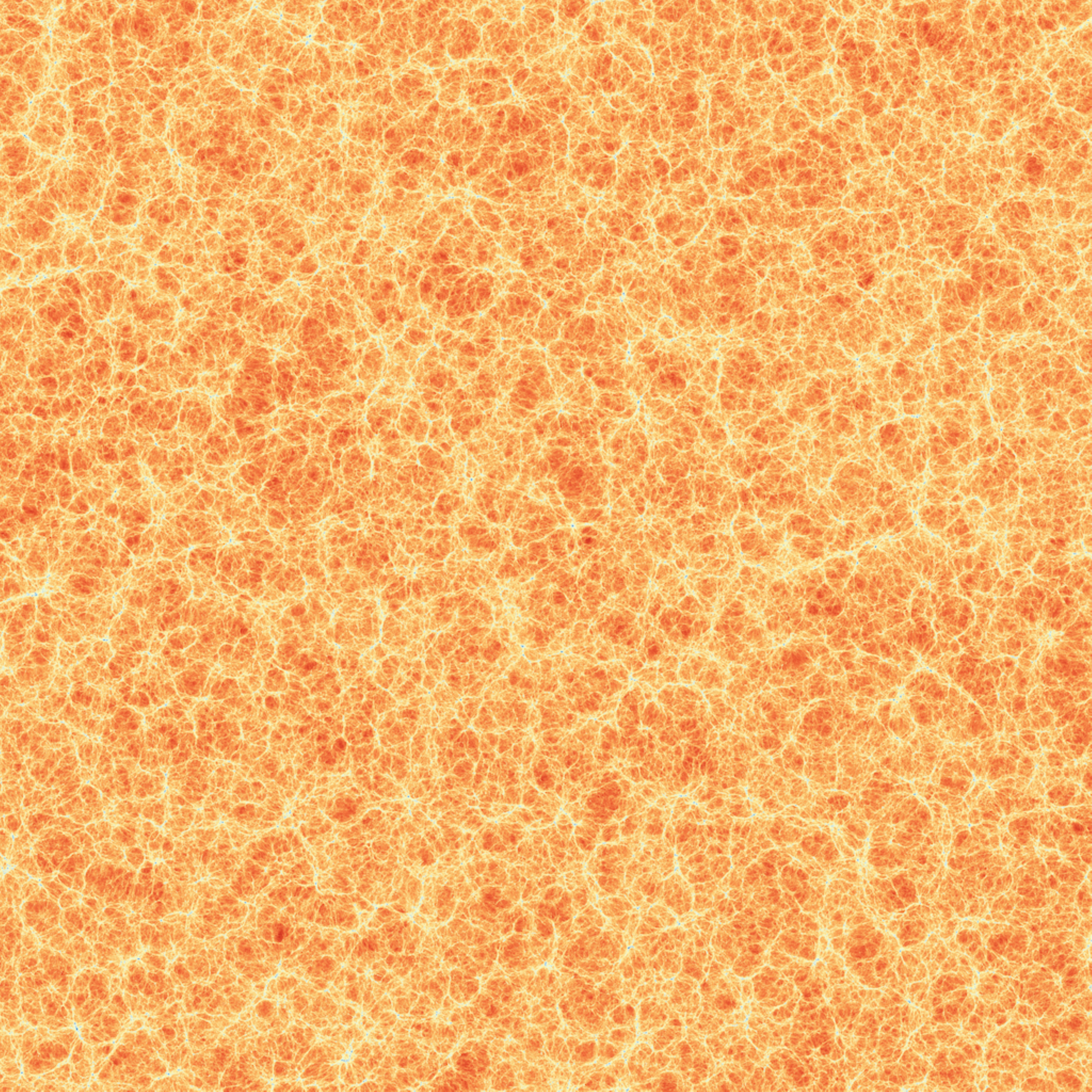}
   \caption{A Slice through the $\Lambda$CDM 500 Mpc$/h$ sampled with $1024^3$ particles dark matter simulation used 
throughout here.}
   \label{fig:simu}
\end{figure}
This paper, makes use of  a dark matter simulation produced with {\tt Gadget2} \citep{gadget2}.
This simulation is characterized by the following $\Lambda$CDM cosmology: $\Omega_{\rm m}=0.265 $, $\Omega_{\Lambda}=0.735$, $n=0.958$, $H_0=70 $ km$\cdot s^{-1} \cdot $Mpc$^{-1}$ and $\sigma _8=0.8$, 
$\Omega_{b}=0.045$
within one standard deviation of WMAP7 results \citep{wmap7}, see Fig.~\ref{fig:simu}.
The box size is 500 Mpc$/h$ sampled with  $1024^3$ particles, the softening length 24 kpc$/h$.
Initial conditions are generated using {\tt mpgrafic}  \citep{mpgrafic}.
The variances and running indexes are measured from the  theoretical  power spectra produced by  {\tt mpgrafic}. 
Snapshots are saved for $z=0, 0.65, 0.97, 1.46, 2.33$ and $3.9$.
An Octree is built for each snapshot, which allows us to count very efficiently all particles 
within a given  sequence of concentric spheres of radii between $R=4,
5 \cdots $ up to $ 21 {\rm Mpc}/h $.  The center of these spheres is sampled regularly on a grid of 
$ 10\, {\rm Mpc}/h $ aside, leading to 117649 estimates of the density per snapshot. All histograms drawn in this paper 
are derived from these samples.  Note that the cells overlap for radii larger than $10 \, {\rm Mpc}/h $.

\section{Mapping cumulants}
\label{App:CumMapping}

In presence of a nonlinear mapping between $\rho$ and $\mu$, such as a logarithmic transformation, cumulants between these two variables follow specific relations that can be computed explicitly in an expansion with respect to the variance. Note that the whole procedure is made complicated because
the notion of connected part is itself dependent on the variable under consideration, so that actually 
$$\langle \rho^n\rangle_c \neq \langle (\exp\mu)^n\rangle_c$$ when the left hand side is computed with respect to the variable $\rho$
and the right hand side to the variable $\mu$. 

The procedure to determine $\langle {\hat\rho}^n\rangle_c$ using the mapping $\rho = \exp \mu$ is the following
\begin{enumerate}
\item Translate the cumulants $\langle\rho^n\rangle_c$ into moments $\langle\rho^n\rangle$ by using the partial Bell polynomials $B_{n,k}$
\begin{align}
\langle {\rho}^n\rangle_c &= \sum_{k=1}^n  (-1)^{k-1} (k-1)! B_{n,k}\left(\{\langle\rho^m\rangle\}_{m=1,\cdots,n}\right) \,. \nonumber
\end{align}
\item Rewrite the moments in terms of $\mu$ using $\langle{\rho}^m\rangle =\langle \exp(m\mu)\rangle =\exp\varphi_\mu(m)$ and normalize the result $\hat\rho = \rho / \langle\rho\rangle$ to enforce $\langle \hat\rho \rangle=1$ 
\begin{align}
\langle{\hat\rho}^n\rangle_c &= \frac{ \sum_{k=1}^n (-1)^{k-1} (k-1)! B_{n,k} \left( \{\exp \varphi_\mu (m) \}_{m=1,\cdots,n}\right)}{[\exp \varphi_\mu(1)]^n}  \,.
\nonumber
\end{align}
\item Finally, use the cumulant expansion theorem to state the result entirely in terms of cumulants $\langle \mu^n\rangle_c$ or reduced cumulants $S_n^\mu= \langle\mu^n\rangle_c / \langle\mu^2\rangle_c^{n-1}$, respectively, 
\begin{align}
\varphi_\mu(m) =\sum_{l=2}^\infty \langle\mu^l\rangle_c \frac{m^l }{l!} =  \sum_{l=2}^\infty S_l^\mu \langle\mu^2\rangle_c^{l-1} \frac{m^l }{l!} \,.
\label{cumexp}
\end{align}
\end{enumerate}
This leads to the following expression for the reduced cumulants where $\sigma_\mu^2 = \langle \mu^2\rangle_c$
\begin{widetext}
\begin{align}
\label{cumrhofrommu}
S_n^\hrho &= \frac{\exp \left[ (n-2) \sum_{l=2}^\infty\limits S_l^\mu \sigma_\mu^{2(l-1)} \frac{1}{l!}\right] 
\sum_{k=1}^n\limits (-1)^{k-1} (k-1)! B_{n,k} \left( \Big\{\exp \Big(  \sum_{l=2}^\infty\limits S_l^\mu \sigma_\mu^{2(l-1)} \frac{m^l }{l!} \Big) \Big\}_{m=1,\cdots,n}\right) }{\left[\exp \Big( \sum_{l=2}^\infty\limits S_l^\mu \sigma_\mu^{2(l-1)} \frac{2^l }{l!} \Big)-\exp \Big( 2\sum_{l=2}^\infty\limits S_l^\mu \sigma_\mu^{2(l-1)} \frac{1}{l!}\Big)\right]^{n-1}} \,.
\end{align}
One can then consistently expand the expression \eqref{cumrhofrommu} up to leading order in $\sigma_\mu^2$. The corresponding results up to the fifth (reduced) cumulant are below, where for simplicity we define $\sigma^2_\mu=\langle\mu^2\rangle_c$. While the cumulants for the density $\rho$ are given in \cite{Fry93}, the reduced cumulants for the normalized density $\hat\rho=\rho/\langle\rho\rangle$ used here read
\begin{align}
 \notag S_3^{\hat\rho}&= (S_3^\mu+3) + \sigma_\mu^2 \left(\frac{3}{2}S_4^\mu+ 2S_3^\mu-2(S_3^\mu)^2+1\right)\,, \\
 S_4^{\hat\rho}&= S_4^\mu+ 12S_3^\mu+16 +\sigma_\mu^2 \left(2S_5^\mu + \frac{45}{2}S_4^\mu -3 S_4^\mu S_3^\mu-18(S_3^\mu)^2 +36S_3^\mu+15\right)\,, \\
 \notag  S_5^{\hat\rho} &= S_5^\mu + 20S_4^\mu+15(S_3^\mu)^2+150S_3^\mu+125  +\sigma_\mu^2 \Bigg(\frac{5}{2}S_6^\mu+48S_5^\mu-4S_3^\mu S_5^\mu+345S_4^\mu+15S_3^\mu S_4^\mu-60(S_3^\mu)^3-60(S_3^\mu)^2+630S_3^\mu+222\Bigg)\,.
\end{align}
\vspace{-0.5cm}
\end{widetext}
\end{document}